\documentclass[useAMS,usenatbib]{mn2e}
\usepackage{threeparttable}
\usepackage{graphicx}
\usepackage{epstopdf}
\usepackage{multirow}
\usepackage{float}

\newcommand\ion[2]{#1$\;${\scshape{#2}}}

\title[The oxygen abundance gradient in M81]{The oxygen abundance gradient in M81
and the robustness of abundance determinations in H~II regions}
\author[]{K.~Z.~Arellano-C\'ordova,\thanks{E-mail: karlaz@inaoep.mx}
M.~Rodr\'iguez, Y.~D.~Mayya, and D.~Rosa-Gonz\'alez\\
Instituto Nacional de Astrof\'isica, \'Optica y Electr\'onica (INAOE),
Apdo. Postal 51 y 216, Puebla, Mexico\\}

\begin{document}

\date{Accepted 2015 October 21. Received 2015 September 25; in original form 2015 April 13}

\maketitle
\label{firstpage}

\begin{abstract}
We study the sensitivity of the methods available for abundance determinations in
\ion{H}{ii} regions to potential observational problems. We compare the dispersions they
introduce around the oxygen and nitrogen abundance gradients when applied to five
different sets of spectra of \ion{H}{ii} regions in the galaxy M81. Our sample contains 116
\ion{H}{ii} regions with galactocentric distances of 3 to 33 kpc, including 48 regions
observed by us with the OSIRIS long-slit spectrograph at the 10.4-m Gran Telescopio Canarias telescope. The
direct method can be applied to 31 regions, where we can get estimates of the electron
temperature. The different methods imply oxygen abundance gradients with slopes of
$-0.010$ to $-0.002$~dex kpc$^{-1}$, and dispersions in the range 0.06--0.25 dex. The
direct method produces the shallowest slope and the largest dispersion, illustrating the
difficulty of obtaining good estimates of the electron temperature. Three of the strong-line
methods, C, ONS, and N2, are remarkably robust, with dispersions of $\sim0.06$ dex, and
slopes in the range $-0.008$ to $-0.006$~dex kpc$^{-1}$. The robustness of each
method can be directly related to its sensitivity to the line intensity ratios that are more
difficult to measure properly. Since the results of the N2 method depend strongly on the N/O
abundance ratio and on the ionization parameter, we recommend the use of the C and ONS
methods when no temperature estimates are available or when they have poor quality,
although the behaviour of these methods when confronted with regions that have different
properties and different values of N/O should be explored.
\end{abstract}


\section{INTRODUCTION}\label{intro}

The analysis of the spectra of \ion{H}{ii} regions provides information about the
chemical composition of the present-day interstellar medium in different kinds of
star-forming galaxies and in different regions across these galaxies. The results
supply fundamental input for our models of galactic chemical evolution. Oxygen, the
third most abundant element, is
taken as representative of the metallicity of the medium, since the oxygen
abundance is the one most easily derived from the optical spectra of photoionized
gas. Leaving aside the construction of photoionization models that reproduce the
spectra, there are different ways to derive oxygen abundances from the observed spectra.
When the spectra are deep enough to allow the measurement of the weak
lines needed for the determination of electron temperatures, such as
[\ion{O}{iii}]~$\lambda4363$ or [\ion{N}{ii}]~$\lambda5755$, we can use the direct
method to derive the O$^+$ and O$^{++}$ abundances, and obtain the total oxygen
abundance by adding these ionic abundances. On the other hand, when the
temperature-sensitive lines are not detected, one must resort to alternative
methods that are based on the intensities of the strongest lines, the so-called
strong-line methods. These methods are calibrated using grids of photoionization
models or samples of \ion{H}{ii} regions that have estimates of the electron
temperature (the empirical methods).

The strongest lines in the optical spectra of \ion{H}{ii} regions that are usually
used by strong-line methods are [\ion{O}{ii}]~$\lambda3727$,
[\ion{O}{iii}]~$\lambda\lambda4959,5007$, [\ion{N}{ii}]~$\lambda\lambda6548+84$,
[\ion{S}{ii}]~$\lambda\lambda6717+31$, H$\alpha$, and H$\beta$. Different methods
use different combinations of line ratios involving these lines and, although a
large variety of methods are available, it is important to consider the procedures
that have been used to calibrate them, and whether the samples of observed objects
or photoionization models used for the calibration cover the same physical
properties as the \ion{H}{ii} regions to which the method will be applied
\citep{Sta:10b}. In general, different methods and different calibrations of the
same method will lead to different results.

It is not easy to construct grids of photoionization models that reproduce well
enough the main characteristics of the observed \ion{H}{ii} regions so that they can be
used to calibrate the strong-line methods \citep{Dop:06,Sta:08}. This
might explain the fact that the abundances derived with methods based on this type of
calibration differ from those derived with empirical methods
\citep{Kew:08a,Lo:10a}. As an example of the complications that arise when
defining the input parameters of photoionization models, we do not have much information
about the properties of dust grains inside \ion{H}{ii} regions \citep[see e.g.][]{Och:15}
and they have important effects on the emitted spectrum \citep{vanH:04}, especially at
high metallicities. Empirical methods also have their problems: it is difficult to
measure the lines needed for temperature determinations in metal-rich regions, the
electron temperatures estimated for these regions can introduce important biases in
the abundance determinations \citep{Sta:05}, and if, as suggested by several
authors, there are temperature fluctuations in \ion{H}{ii} regions which are larger than
those predicted by photoionization models, they can lead to lower abundances than the
real ones at any metallicity \citep[e.g.][]{Pena:12a}.

If one excludes from the samples high-metallicity objects, and if temperature
fluctuations turn out to be not much higher than the ones expected from
photoionization models, it can be argued that empirical calibrations of the
strong-line methods should be preferred because they are based on a lower number or
assumptions, although photoionization models can provide much insight on the
explanations behind the behaviour and applicability of the strong-line methods. One
important question is how well strong-line methods can be expected to do. Grids of photoionization models can be used to show that strong-line methods work because
the metallicity of most \ion{H}{ii} regions is strongly related to the effective
temperature of the ionizing radiation and to the ionization parameter of the
region\footnote{The number of ionizing photons per atom arriving to the inner face of the
ionized region.}\citep{Dop:06,Sta:08}. This implies that strong-line methods will
not work properly when applied to regions that do not follow this general relation
due to variations in their star formation histories, ages, or chemical evolution
histories \citep{Sta:10b}. The direct method is expected
to work better since it is based on a smaller number of assumptions, and when
observations of \ion{H}{ii} regions are presented in any publication, it is usually
described as an achievement to detect the weak lines that allow a temperature
determination.

However, the measurement of the weak, temperature-sensitive, lines can be affected
by large uncertainties when these lines have a low signal-to-noise ratio in the
nebular spectrum. When the oxygen abundances are derived with the direct method
using temperature estimates based on these lines, the results will also have large
uncertainties. The calibration of strong-line methods using these oxygen abundances
can be affected by the large uncertainties, but this problem can be alleviated by a
careful selection of calibration samples trying to have small, randomly
distributed, uncertainties, and by cleaning up the samples excluding the outliers,
since it can be assumed that they depart from the relation implied by the
rest of the sample either because they have different properties or because their
line intensities have large uncertainties.
In principle the average behaviour of these samples could allow good calibrations
of strong-line methods which might then show lower dispersions than the results of
the direct method when applied to objects in the calibration sample or to objects
that have the average properties of the calibration sample. In these cases,
strong-line methods will be more robust than the direct method.

The measurement of the intensities of the strong lines used by the strong-line
methods should present less problems. However, there are observational effects that
introduce uncertainties in all the measurements of line intensity ratios, effects
that are not necessarily included in the estimated uncertainties, namely,
atmospheric differential refraction leading to the measurement of different lines
at different spatial positions, the incorrect extraction of 1D spectra from tilted
2D spectra, undetected absorption features beneath the emission lines, problems
with the estimation of the continuum or with deblending procedures, the presence of
unnoticed cosmic rays, or any bias introduced by the flux calibration or the
extinction correction. Some of the line ratios used by strong-line methods will be
more sensitive to these effects, making these methods less robust than others that
are based on less-sensitive line ratios. Moreover, since the line ratios used
as temperature diagnostics can be very sensitive to these observational problems, the
results of the direct method might be less robust than those derived with strong-line
methods even when the weak temperature-sensitive lines are measured with a good
signal-to-noise ratio.

One way to infer the robustness of the methods used for abundance determinations in
the presence of observational problems is to compare their performance when they
are used to estimate metallicity gradients in galaxies. The observational problems
are likely to introduce dispersions around an existing gradient that can be
interpreted as azimuthal abundance variations. If any of the methods implies
significantly lower dispersions, it seems reasonable to assume that azimuthal
variations must be lower than the estimated dispersions, and hence that the method
is behaving in a more robust way. Since spectra obtained by different authors are
likely to be affected by various observational problems in different amounts, the
robustness of each method to observational effects can be inferred from the
dispersions around the gradient implied by the method when using spectra observed
by different authors in the same galaxy. Methods that show significantly lower
dispersions can then be inferred to be more robust.

Here we present an analysis of the oxygen abundance gradient in M81, using this
galaxy as a case study of the robustness of some of the methods used for abundance
determinations in \ion{H}{ii} regions. We will explore the behaviour of methods
that have been calibrated using large samples of \ion{H}{ii} regions that have
temperature determinations. M81 is an ideal candidate for this study, since it is a
nearby spiral galaxy, at a distance of 3.63$\pm$0.34 Mpc~\citep{Free:01a}. This
galaxy belongs to an interacting group of galaxies and has well-defined spiral
arms that contain a large number of \ion{H}{ii} regions. The oxygen abundance
gradient of M81 has been calculated in different studies using several methods
(\citealt{Stau:84a,Gar:87a,Pil:04a,Stan:10a,Patt:12a,Stan:14a};
\citealp*{Pil:14a}). These works find slopes that go from $-0.093$ to $-0.011$ dex
kpc$^{-1}$, and some of them include \ion{H}{ii} regions where it is possible
to measure the electron temperature and calculate the metallicity with the direct method.

This paper is structured as follows: in Section~2 we describe our observations,
which were obtained with the Gran Teles\-copio Canarias (GTC), the data reduction, the
sample selection, the measurement of the line intensities, and the reddening corrections;
in Section~3 we describe the methods we apply to calculate the physical conditions and
chemical abundances of the sample of \ion{H}{ii} regions; in Section 4 we present
the results of this analysis, and the implied metallicity gradients, using our data and
other observations from the literature; in Section 5 we discuss the
scatter around the metallicity gradient implied by the different methods; and finally,
in Section~6, we summarize our results and present our conclusions. 


\section[Observaciones]{OBSERVATIONS AND DATA REDUCTION}

Spectroscopic observations (programme GTC11-10AMEX, PI: DRG) were carried out using the
long-slit spectrograph of the OSIRIS instrument at the 10.4-m GTC telescope in the
Observatorio del Roque de los Muchachos (La Palma, Spain). We used the five slit
positions listed in Table~\ref{Slits}, with a slit width of 1 arcsec and length of
8 arcmin. Table~\ref{Slits} provides the central positions of the slits, the exposure
times we used, the slit position angles (P.A.), and the airmasses during the
observations. We obtained three exposures of 900 s
at each slit position using the R1000B grism, which allowed us to cover the
spectral range 3630--7500 \AA\ with a spectral resolution of $\sim7$ \AA\ full width at
half-maximun.
The observations were acquired on 2010 April 5--7 when the seeing was $\sim1$ arcsec.
The detector binning by 2 pixels in the spatial dimension provided a scale of 0.25 arcsec
pixel$^{-1}$. 
The airmasses were in the range 1.3--1.5 and, at these values, departures from the
parallactic angle can introduce light losses at some wavelengths due to differential
atmospheric refraction~\citep{Fil:82}. In our observations, the differences between
the position angle and the parallactic angle go from 8 to 23 degrees. Although small,
the differences imply that we might be losing some light in the blue, especially for
the few objects with sizes around 1 arcsec observed with slit positions P1 and P2.
This is one of the possible observational problems that we listed in Section~\ref{intro}, 
and the combined effects of these problems are explored in our analysis.

The data were reduced using
the tasks available in the \textsc{iraf}\footnote{\textsc{iraf} is distributed by the National
Optical Astronomy Observatory, which is operated by the Association of Universities
for Research in Astronomy, Inc., under cooperative agreement with the National
Science Foundation.} software package. The reduction process included bias
subtraction, flat-field and illumination correction, sky subtraction, wavelength
calibration, and flux calibration using the standard star Feige 34. The final
spectra result from the median of the three exposures obtained at each slit
position.

\begin{table}
\caption{Log of the observations.}
\begin{tabular}{lcccrc}
\hline
Slit & R.A. & Dec. & Exposure & P.A.& Airmass\\
ID & (J2000) & (J2000) & times (s)& ($\degr$)\\
\hline
P1& 09:54:38 & +69:05:48 & $3\times900$ & 171 & 1.3\\ 
P2& 09:54:52 & +69:08:11 & $3\times900$ & 6      & 1.3\\
P3& 09:55:37 & +69:07:46 & $3\times900$ & 123 &  1.4\\
P4& 09:55:46 & +69:07:48 & $3\times900$ & 105 &  1.5\\
P5& 09:55:48 & +69:04:53 & $3\times900$ & 127 &   1.4\\ 
\hline
\end{tabular}
\label{Slits}
\end{table}

The slit positions were selected to pass through some of the brightest stellar
compact clusters in the catalogue of \citet*{Santi:10a} for M81. These observations
are part of a large-scale program dedicated to study the star formation in this
galaxy \citep{Mayya:13}. Here we use them to study the chemical abundances and the
abundance gradient provided by \ion{H}{ii} regions in M81. We extracted spectra
using the task \textsc{apall} of \textsc{iraf} for each knot of ionized gas that
we found along the five slits. There were two or three bright stellar clusters in
each slit and we used the one closest to each ionized knot to trace the small changes
of position of the stellar continuum in the CCD. We fitted polynomial functions to
these traces and used them as a reference to extract the spectrum of the knots.
The size of the apertures goes from 4 to 28 pixels (1 to 7 arcsec). The
final sample consists of 48 \ion{H}{ii} regions located in the disc of M81.

\begin{figure*}
\begin{center}
\includegraphics[width=0.99\textwidth]{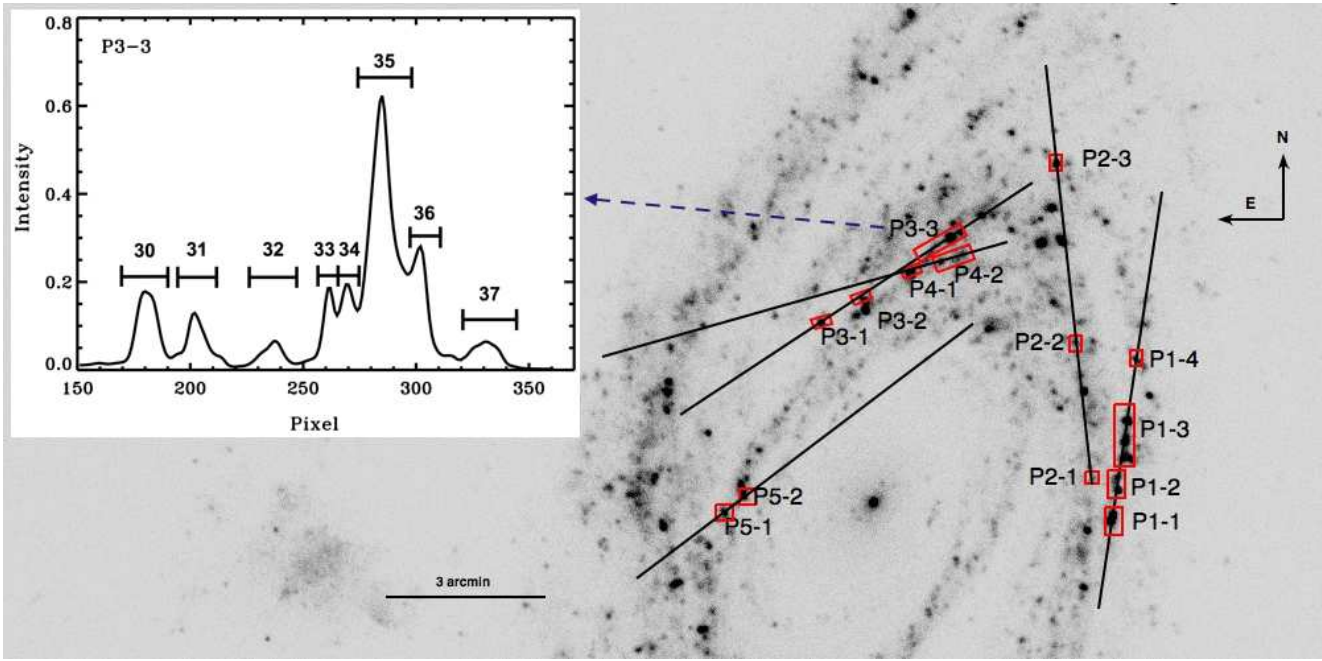}
\caption{UV image of M81 from \textit{GALEX} showing the slit positions listed in
Table~\ref{Slits}. The boxes show the locations of the ionized knots in our sample.
The inset shows a cut in the spatial direction along the H$\alpha$ emission line
for box P3-3. We identify in the inset the knots of ionized gas whose spectra we
extracted in this region.} 
\label{slit-regions}
\end{center}
\end{figure*}

\begin{figure*}
\begin{center}
\includegraphics[width=0.8\textwidth]{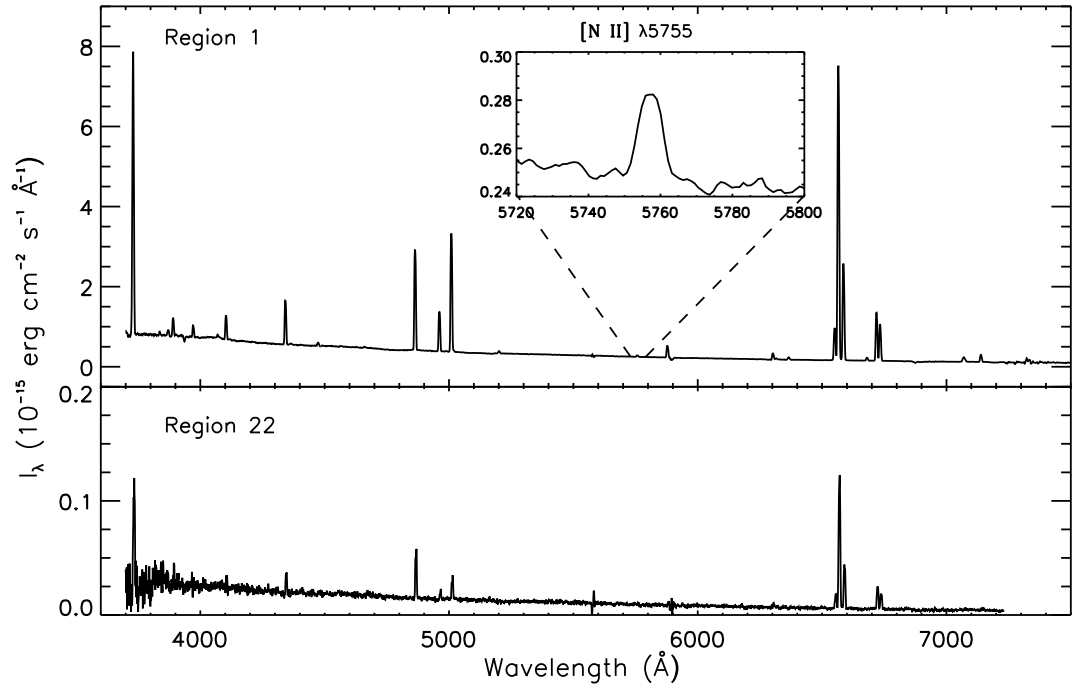}
\caption{Spectra of two of our observed regions. Region~1 has one of the spectra
with the highest signal-to-noise ratios; region~22 has one of the lowest signal-to-noise 
ratios. The inset shows our detection of the temperature-sensitive line
[\ion{N}{ii}] $\lambda5755$ in region~1.} 
\label{spectra}
\end{center}
\end{figure*}

Fig.~\ref{slit-regions} shows the UV image of M81 from \textit{GALEX} (Galaxy Evolution
Explorer) with our slit positions superposed. We also show boxes around the regions
where we could extract spectra for several  knots of ionized gas. One to eight knots were extracted in each of the boxes shown in Fig.~\ref{slit-regions}. The boxes are
tagged as P$n$-$m$, where $n$ identifies the slit and $m$ the box along this slit.
We identify the knots with numbers going from 1 to 48, starting with the first knot
in box P1-1 and ending with the knots in P5-2, moving from from South to North in
the slits P1 and P2 and from East to West for the slits P3, P4, and P5. We also
show an inset in Fig.~\ref{slit-regions} with a cut in the spatial direction along
one of the columns with H$\alpha$ emission in our 2D spectra for box P3-3,
illustrating the procedure we followed for selecting the ionized knots.
Fig.~\ref{spectra} shows two examples of the extracted spectra, one with a high
signal-to-noise ratio (region~1) and a second one with a low signal-to-noise ratio
(region~22).

\subsection{Line measurements}
Line intensities were measured using the \textsc{splot} routine of  \textsc{iraf} by integrating
the flux above the continuum defined by two points on each side of the emission
lines. We fitted Gaussian profiles for those lines that appear blended. The
errors in the line intensities were calculated using the expression
\citep{Tres:99a}:
\begin{equation}
\sigma_{I}=\sigma_{c}D\sqrt{2N_{pix}+\frac{EW}{D}},
\end{equation}
where $D$ is the spectral dispersion in \AA\ per pixel, $\sigma_c$ is the mean
standard deviation per pixel of the continuum on each side of the line, $N_{pix}$
is the number of pixels covered by the line and EW is the equivalent width.
We corrected the Balmer line intensities for the effects of stellar absorption
by assuming absorption equivalent widths of 2~\AA\ \citep{Mc:85}. The correction is
small for most of our regions, with changes in the H$\alpha$ and H$\beta$ intensities
below 7 and 10 per cent, respectively, but it is significant for six regions. In four of
them (regions 7, 42, 43, and 44) it increases the intensity of H$\beta$ by just 12--14
per cent, but regions 6 and 14 have increments of 72 and 27 per cent, respectively.
The effects of these changes on our results are described in Section~\ref{Oab}.

The emission lines were corrected for extinction assuming an intrinsic line ratio of
H$\alpha$/H$\beta=2.86$, suitable for $T_{\rm e}=10000$ K and $n_{\rm e}=100$
cm$^{-3}$ \citep{Os:06a}, since we find similar values for the physical
conditions in our objects. We used the extinction law of \citet*{Car:89a} with a
ratio of total to selective extinction in $V$ and $B-V$ of $R_V=3.1$.

To correct for reddening each emission line ratio, we use the expression:
\begin{equation}
\frac{I({\lambda})}{I({\mbox{H}\beta})}=
\frac{I_0({\lambda)}}{I_0(\mbox{H}\beta)}10^{-c({\rm H}\beta)[f(\lambda)-1]}
\end{equation} 
where $I(\lambda)/I(\mbox{H}\beta)$ is the observed line intensity ratio,
$I_0(\lambda)/I_0(\mbox{H}\beta)$ is the reddening-corrected ratio,
$c(\mbox{H}\beta)$ is the reddening coefficient, and $f(\lambda)$ is the
extinction law normalized to H$\beta$.

Tables~\ref{chb} and \ref{lines}, whose full versions are available online,
show the values of the extinction coefficients and the observed and
reddening-corrected line ratios for each region. We also provide for each region
the extinction corrected $I(\mbox{H}\beta)$ in Table~\ref{chb}. The final errors
are the result of adding quadratically the uncertainties in the measured intensities,
4 per cent as our estimate of the uncertainty in the flux calibration, and the
uncertainty in the reddening correction. The values we
find for $c(\mbox{H}\beta)$ are in the range 0--0.51, in agreement with the
values found by \citet{Patt:12a} for several \ion{H}{ii} regions in M81,
$c(\mbox{H}\beta)=0.07\mbox{--}0.43$ but significantly lower than the values obtained
by \citet{Stan:10a} for \ion{H}{ii} regions in this galaxy,
$c(\mbox{H}\beta)=0.48\mbox{--}0.92$.

\begin{table}
\caption{The extinction coefficients $c(\mbox{H}\beta)$ and the reddening-corrected intensities
for H$\beta$. The full table for the 48 regions is available online.}
\label{chb}
\begin{tabular}{cccc}
\hline
\multicolumn{1}{c}{Region} & \multicolumn{1}{c}{$c(\mbox{H}\beta)$}& \multicolumn{1}{c}{Error}  & \multicolumn{1}{c}{ $I(H\beta$)}  \\
 & & & (erg cm$^{-2}$ s$^{-1}$)\\ 
\hline
 1   &   0.14 & 0.07  &    2.75$\times10^{-14}$   \\   
 2   &   0.35 & 0.08  &    5.95$\times10^{-15}$  \\   
 3   &   0.39 & 0.07  &    8.20$\times10^{-15}$   \\   
 4   &   0.00 & 0.10  &    5.51$\times10^{-16}$   \\   
 5   &   0.06 & 0.08  &    1.37$\times10^{-15}$   \\   
 6   &   0.00 & 0.09  &    2.84$\times10^{-16}$   \\   
 7   &   0.28 & 0.08  &    7.87$\times10^{-16}$   \\   
 8   &   0.00 & 0.07  &    3.60$\times10^{-15}$   \\   
 9   &   0.25 & 0.09  &    9.44$\times10^{-16}$   \\   
10   &   0.35 & 0.08  &    1.25$\times10^{-14}$   \\     
\hline
\end{tabular}
\end{table}

\begin{table}
\caption{Some of the observed and reddening-corrected line ratios, normalized to
$I(\mbox{H}\beta)=100$, for region~1. The error is expressed as a percentage
of the reddening-corrected values. The full table with the line
intensities for the 48 regions is available online.}
\label{lines}
\begin{tabular}{cllrrcc}
\hline
\multicolumn{1}{l}{Region} & \multicolumn{1}{l}{$\lambda$(\AA)} & ID & \multicolumn{1}{r}{$I(\lambda)$}
& \multicolumn{1}{r}{$I_0(\lambda$)} &  \multicolumn{1}{c}{Error (\%)} \\
\hline
1 & 3727  &   $[$\ion{O}{ii}$]$    &    266    &   306      &  8  \\
1 &4101  &   H$\delta$            &    21.4   &   23.7     &  7  \\
1 &4341  &   H$\gamma$            &    41.2   &   44.2     &  6  \\
1 &4471  &   \ion{He}{i}          &    3.3    &   3.5      &  8  \\
1 &4861  &   H$\beta$             &    100.0  &   100.0    &  5  \\
1 &4959  &   $[$\ion{O}{iii}$]$   &    40.0   &   39.5     &  5  \\
1 &5007  &   $[$\ion{O}{iii}$]$   &    120.1  &   118.1    &  5  \\
1 &5200  &   $[$\ion{N}{i}$]$     &    2.5    &   2.5      &  8  \\
1 &5755  &   $[$\ion{N}{ii}$]$    &    1.4    &   1.3      &  9  \\
1 &5876  &   \ion{He}{i}          &    11.8   &   10.8     &  6  \\
\hline
\end{tabular}
\end{table}


\section{Physical conditions and oxygen abundances}

\subsection{The direct method}

We could measure the temperature-sensitive [\ion{N}{ii}] $\lambda5755$ line in 12
of the 48 \ion{H}{ii} regions in our sample, where it
shows a well-defined profile with a S/N $\geq$ 3.6 (see e.g. Fig.~\ref{spectra}).
This allows us to use the so-called direct method to derive the oxygen abundances,
which is, in principle, the most reliable method. The [\ion{O}{iii}]~$\lambda4363$
auroral line was marginally detected in two regions with a noisy profile.
The line can be affected by imperfect sky subtraction of the
Hg~$\lambda4358$ sky line, and we decided not to use it.

In order to calculate the physical conditions and the ionic oxygen abundances in
these 12 \ion{H}{ii} regions, we use the tasks available in the \textsc{nebular} package
of \textsc{iraf}, originally based on the calculations of \citet*{Rob:87a} and
\citet{Shaw:95a}. We adopted the following atomic data: the transition
probabilities of \citet{Ze:82a} for O$^+$, \citet*{wi:96a} and \citet{Sto:00a} for
O$^{++}$, \citet{wi:96a} for N$^+$ and \citet{Men:82a} for S$^+$; and the
effective collision strengths of \citet{Prad:06a} for O$^+$, \citet{agg:99a} for
O$^{++}$, \citet{Len:94a} for N$^+$, and \citet{Kee:96a} for S$^+$. 

We use the line intensity ratio [\ion{S}{ii}] $\lambda6717/\lambda6731$ to
calculate the electron density, $n_{\rm e}$, and [\ion{N}{ii}]
$(\lambda6548+\lambda6583)/\lambda5755$  to calculate the electron temperature. The
[\ion{S}{ii}] ratio could be measured in all the regions, and we used $T_{\rm
e}=10000$ K to derive $n_{\rm e}$ in those regions where the [\ion{N}{ii}]
$\lambda5755$ line was not available. We obtain $n_{\rm e}\la100$ cm$^{-3}$ in most
of the regions. At these densities, the [\ion{S}{ii}] diagnostic is not very
sensitive to density variations \citep{Os:06a} and, in fact, some of the regions have
a line ratio that lies above the range of expected values. However, all the
[\ion{S}{ii}] line ratios but two are consistent within one sigma with $n_{\rm
e}\la100$ cm$^{-3}$ and, since for these values of $n_{\rm e}$ the derived ionic
abundances show a slight dependence on density, we use $n_{\rm e}=100$ cm$^{-3}$ in
all our calculations. On the other hand, the upper level of the [N II] $\lambda5755$
line can be populated by transitions resulting from recombination, leading to an
overestimate of the electron temperature \citep{Rub:86}. We used the expression
derived by \citet{Liu:00a} to estimate a correction for this contribution, but found
that the effect is very small in our objects, $\la40$ K in $T_{\rm e}$, so that it is
safe to ignore this correction. The values derived for $n_{\rm e}$ and $T_{\rm
e}$([\ion{N}{ii}]) are listed in Table~\ref{Oxygen-abundances}, where we use `:' to
identify the most uncertain values of $n_{\rm e}$. Table~\ref{Oxygen-abundances} also
gives for all the objects in our sample the number that we use for identification
purposes, the coordinates of the region, the slit and box where the spectra were
extracted, the angular sizes of the extracted regions, their galactocentric distances
(see Section~\ref{Ograd}), and their oxygen and nitrogen abundances derived with
the methods described below.

We adopt a two-zone ionization structure characterized by
$T_{\rm e}$([\ion{N}{ii}]) in the [\ion{O}{ii}] emitting region and by
$T_{\rm e}$([\ion{O}{iii}]) in the [\ion{O}{iii}] emitting region, where the value
of $T_{\rm e}$([\ion{O}{iii}]) is obtained using the relation given by
\citet[see also \citealp{Gar:92a}]{Camp:86a}: 
\begin{equation}
T_{\rm e}([\mbox{\ion{N}{ii}}])\simeq T_{\rm e}([\mbox{\ion{O}{ii}}])=
0.7\, T_{\rm e}([\mbox{\ion{O}{iii}}]) + 3000\ \mbox{K},
\label{Relation}
\end{equation}
which is based on the photoionization models of \citet{Sta:82}. This relation is
widely used (see e.g. \citealt{Bre:11a, Patt:12a, Pil:12a}) and is similar to the
one obtained from good-quality observations of \ion{H}{ii} regions \citep{Est:09a}.

The ionic oxygen abundances are derived using the physical conditions described
above and the intensities of [\ion{O}{ii}]~$\lambda3727$ and
[\ion{O}{iii}]~$\lambda\lambda4959, 5007$ with respect to H$\beta$. The final
values of the oxygen abundances can be obtained by adding the contribution of both
ions: $\mbox{O}/\mbox{H}=\mbox{O}^+/\mbox{H}^++\mbox{O}^{++}/\mbox{H}^+$.
The N abundance is calculated using the [\ion{N}{ii}]~$\lambda\lambda6548+84$
lines and the assumption that N/O~$\simeq\mbox{N}^+/\mbox{O}^+$.


\subsection{Strong-line methods}

When the emission lines needed to derive the electron temperature are too weak to
be observed, it is still possible to estimate chemical abundances with the
so-called strong-line methods. These methods are based on the intensities of lines
that can be easily measured, such as [\ion{O}{ii}] $\lambda3727$, [\ion{O}{iii}]
$\lambda5007$, or [\ion{N}{ii}] $\lambda6584$, and are calibrated using
photoionization models or observational data of \ion{H}{ii} regions that include
measurements of the electron temperature. The two approaches often lead to
different results \citep[see e.g.,][]{Kew:08a}, but we will not enter here into a
discussion of which one yields the better estimates; we will use the empirical
methods just because they provide the simplest approach to the problem. We have
selected some of the empirical calibrations that are based on the largest numbers
of \ion{H}{ii} regions: the P method of \citet{Pil:05a}, the ONS method
of \citet{Pil:10a}, the C method of \citet{Pil:12a} and the O3N2 and N2 methods
calibrated by \citet{Marino:13a}. The methods use initial samples of around
100--700 \ion{H}{ii} regions that have temperature measurements, although in some
cases different criteria are applied in order to select more adequate or more
reliable subsamples. All these methods provide estimates of the oxygen abundance,
whereas nitrogen abundances can only be obtained with the ONS and C methods.
We describe the methods below.

\subsubsection{The P method}

Some of the most widely used strong-line methods are based on the parameter
$R_{23}=I([$\ion{O}{ii}$]~\lambda3727)/I(\mbox{H}\beta)+
I([$\ion{O}{iii}$]~\lambda\lambda4959,5007)/I(\mbox{H}\beta)$, first introduced
by~\citet{Pa:79a}. There are many different calibrations of this method, and they
can lead to oxygen abundances up to 0.5 dex above those obtained from the direct
method \citep*{Kenni:03a}. Here we use the calibration of \citet{Pil:05a}, which is
based on a large sample of \ion{H}{ii} regions that have temperature measurements.
This calibration is called the P method because it uses as a second parameter in
the abundance determination an estimate of the hardness of the ionizing radiation,
$P=I([$\ion{O}{iii}$]~\lambda\lambda4959,5007)/(I([$\ion{O}{iii}$]~\lambda\lambda4959,5007)
+I([$\ion{O}{ii}$]~\lambda3727))$, as
proposed by \citet{Pil:01a,Pil:01b}. According to \citet{Pil:05a}, this method
provides oxygen abundances that differ by less than 0.1 dex from the values
obtained with the direct method.

The main problem with the methods based on $R_{23}$ is that the relation of this
parameter with $12+\log(\mbox{O}/\mbox{H})$ is double valued: the same value of
$R_{23}$ can lead to two different values of the oxygen abundance and one must find
a procedure to break this degeneracy. Following \citet{Kew:08a}, we use
$\log(I([$\ion{N}{ii}$]~\lambda6584)/I([$\ion{O}{ii}$]~\lambda3727))=-1.2$ as the
dividing line between low- and high-metallicity objects.

\subsubsection{The ONS method}

The ONS method, proposed by \citet{Pil:10a}, uses the relative intensities of the
lines [\ion{O}{ii}]~$\lambda3727$, [\ion{O}{iii}]~$\lambda\lambda4959,5007$,
[\ion{N}{ii}]~$\lambda6548+84$, [\ion{S}{ii}]~$\lambda6717+31$, and H$\beta$.
\citet{Pil:10a} classify the \ion{H}{ii} regions as cool, warm, or hot depending
on the relative intensities of the [\ion{N}{ii}], [\ion{S}{ii}] and H$\beta$ lines,
and provide different formulae that relate the oxygen and nitrogen abundances
to several line ratios for each case.  \citet{Pil:10a} find that the method shows very
good agreement with the abundances they derive using the direct method, with root mean
square differences of 0.075 dex for the oxygen abundance and 0.05 dex for the
nitrogen abundance. \citet*{Li:13a} find similar differences with the direct
method for their sample \ion{H}{ii} regions, around 0.09 dex in the oxygen abundance.

\subsubsection{The C method}

The counterpart method or C method of \citet{Pil:12a} is based on the assumption
that \ion{H}{ii} regions that have similar intensities in their strong emission
lines have similar physical properties and chemical abundances. The method uses a
data base of 414 reference \ion{H}{ii} regions that are considered to have good
estimates of the electron temperature, and looks for objects that have values which
are similar to the ones observed in the \ion{H}{ii} region under study for several
line ratios involving the lines [\ion{O}{ii}]~$\lambda3727$,
[\ion{O}{iii}]~$\lambda5007$, [\ion{N}{ii}]~$\lambda6584$,
[\ion{S}{ii}]~$\lambda6717+31$, and H$\beta$. The method then finds a relation
between the oxygen or nitrogen abundance and the values of these line intensity
ratios for these objects, which is then applied to derive the oxygen abundance of the observed
\ion{H}{ii} region. \citet{Pil:12a} estimate that if the errors in the line
intensity ratios are below 10 per cent, the method leads to abundance uncertainties
ofless than 0.1 dex in the oxygen abundance, and 0.15 dex in the nitrogen abundance.

\subsubsection{The O3N2 and N2 methods}

The O3N2 and N2 methods were proposed by \citet{All:79a} and \citet{Sto:94}, respectively. 
They use the line ratios:
\begin{equation}
\mbox{O3N2}=\log\left(\frac{I([\mbox{\ion{O}{iii}}]~\lambda5007)/I(\mbox{H}\beta)}
{I([\mbox{\ion{N}{ii}}]~\lambda6584)/I(\mbox{H}\alpha)}\right)
\end{equation}
and 
\begin{equation}
\mbox{N2}=\log(I([\mbox{\ion{N}{ii}}]~\lambda6584)/I(\mbox{H}\alpha)).
\end{equation}
These methods are not sensitive to the extinction correction or flux calibration
and have been widely used. However, the O3N2 method cannot be used at low
metallicities, the N2 method can be affected by shocks or the presence of an AGN
in nuclear H II regions
\citep{Kew:02a}, and both methods are very sensitive to the degree of ionization of
the observed region and to its value of N/O. This might explain the large
dispersions usually found in their calibration, although this could also be due to
the selection of the calibration sample. We will use the calibrations of
\citet{Marino:13a} for these two methods that are based on \ion{H}{ii} regions
with temperature measurements. The root mean square differences between the oxygen
abundances derived with these methods and those derived with the direct method for
the objects used by \citet{Marino:13a} are 0.16 dex (N2 method) and 0.18 dex
(O3N2 method).


\section{Results}\label{Ograd}

\subsection{Oxygen abundances and the oxygen abundance gradient}\label{Oab}

Table~\ref{Oxygen-abundances} shows the oxygen abundances derived for the 48
regions in our sample using the methods described above. The uncertainties provided
for the results of the direct method are those arising from the estimated errors in
the line intensities. For the results of the P and ONS methods, we have added
quadratically the estimated uncertainties of the methods, 0.1 dex, to the
uncertainties in the measured line ratios. In the case of the ONS method, the
derived uncertainties are in the range 0.10--0.12 dex in all cases, and we decided
to adopt an uncertainty of 0.12 dex for this method. For the C method we adopt an
uncertainty of 0.10 dex, the value estimated by \citet{Pil:12a} for the case when
the line ratios involved in the calculations have uncertainties below 10 per cent.
Some of the regions have line ratios with larger uncertainties, up to 40 per cent,
but our results below agree with uncertainties around or below 0.10 dex for the
oxygen abundances derived with this method in most of the \ion{H}{ii} regions. We
assigned uncertainties of 0.16 and 0.18 dex for the N2 and O3N2 methods,
respectively, the ones found in the calibration of these methods, since the errors
in the line intensities do not add significantly to this result.

\begin{table*}
\begin{minipage}{170mm}
\caption{Coordinates, sizes, galactocentric distances, physical conditions and oxygen
abundances for the 48 \ion{H}{ii} regions in our sample. The oxygen abundances
have been derived with the direct method ($T_{\rm e}$) and five
strong-line methods (P, ONS, C, O3N2, and N2).}
\begin{tabular}{lccccccccccccc}
\hline
\multicolumn{1}{l}{ID} & \multicolumn{1}{c}{Box} & \multicolumn{1}{c}{RA} & \multicolumn{1}{c}{Dec.} & \multicolumn{1}{c}{Size} &
\multicolumn{1}{c}{$R$} & \multicolumn{1}{c}{$n_{\rm e}$} &
\multicolumn{1}{c}{$T_{\rm e}$([\ion{N}{ii}])} &
\multicolumn{6}{c}{$12+\log(\mbox{O}/\mbox{H})$} \\
 & & (J2000)& (J2000) & (arcsec) & (kpc) & (cm$^{-3}$) & (K) & ($T_{\rm e}$) & (P) & (ONS) &
 (C) & (O3N2) & (N2) \\ 
\hline
1  & P1-1   &   09:54:43   &   +69:03:39   &        5.1 & 8.9 & 115$\pm$6   & 10100$^{+500}_{-400}$   & $8.13^{+0.06}_{-0.07}$  &   8.33 &  8.45 & 8.43 & 8.42  & 8.53 \\
2  &        &   09:54:43   &   +69:03:33   &        2.1 & 8.9 & $-$        &         & $8.09^{+0.08}_{-0.09}$  &   8.22 &  8.45 & 8.52 & 8.45  & 8.54 \\
3  &        &   09:54:43   &   +69:03:31   &        2.5 & 8.9 & 27$\pm$21        & $-$                     & $-$                     &   8.42 &  8.48 & 8.49 & 8.39  & 8.59 \\
4  &        &   09:54:43   &   +69:03:24   &        2.0 & 9.0 & $-$        & $-$                     & $-$                     &   8.08 &  8.40 & 8.44 & 8.43  & 8.52 \\
5  & P1-2   &   09:54:41   &   +69:04:23   &        4.6 & 8.7 & $-$        & 10900$^{+1700}_{-1000}$ & 7.96$^{+0.16}_{-0.20}$  &   8.22 &  8.46 & 8.55 & 8.49  & 8.56 \\
6  &        &   09:54:42   &   +69:04:08   &        1.5 & 8.8 & 314:       & $-$                     & $-$                     &   8.53 &  8.52 & 8.47 & 8.41  & 8.48 \\
7  &        &   09:54:42   &   +69:04:06   &        2.6 & 8.8 & $-$        & $-$                     & $-$                     &   8.30 &  8.46 & 8.49 & 8.43  & 8.50 \\
8  & P1-3   &   09:54:39   &   +69:05:01   &        2.9 & 8.7 & 23:        & 9400$^{+800}_{-600}$    & 8.17$^{+0.11}_{-0.12}$  &   8.43 &  8.49 & 8.46 & 8.45  & 8.53 \\
9  &        &   09:54:40   &   +69:04:58   &        2.0 & 8.7 & $-$        & $-$                     & $-$                     &   8.27 &  8.57 & 8.57 & 8.58  & 8.54 \\
10 &        &   09:54:40   &   +69:04:49   &        6.2 & 8.7 & 84$\pm$22   & 10400$^{+500}_{-400}$   & 8.13$\pm$0.07           &   8.33 &  8.44 & 8.40 & 8.40  & 8.55 \\
11 &        &   09:54:40   &   +69:04:40   &        5.1 & 8.7 & 19:        & $-$                     & $-$                     &   8.29 &  8.50 & 8.50 & 8.52  & 8.57 \\
12 &        &   09:54:40   &   +69:05:06   &        3.7 & 8.7 & $-$        & 10600$^{+1800}_{-1000}$ & 8.08$^{+0.16}_{-0.20}$  &   8.21 &  8.43 & 8.51 & 8.43  & 8.53 \\
13 &        &   09:54:40   &   +69:05:10   &        1.0 & 8.7 & 2:         & $-$                     & $-$                     &   8.24 &  8.59 & 8.56 & 8.58  & 8.54 \\
14 &        &   09:54:40   &   +69:05:13   &        2.3 & 8.7 & 50:        & $-$                     & $-$                     &   8.06 &  8.51 & 8.53 & 8.58  & 8.59 \\
15 &        &   09:54:40   &   +69:05:24   &        5.4 & 8.7 & 3:         & $-$                     & $-$                     &   8.28 &  8.58 & 8.56 & 8.57  & 8.52 \\
16 & P1-4   &   09:54:38   &   +69:06:38   &        4.9 & 8.5 & $-$        & $-$                     & $-$                     &   8.44 &  8.53 & 8.52 & 8.53  & 8.56 \\
17 & P2-1   &   09:54:47   &   +69:04:25   &        3.1 & 7.7 & $-$       & 13300$^{+1800}_{-1200}$ & 7.70$^{+0.11}_{-0.12}$  &   8.51 &  8.50 & 8.34 & 8.27  & 8.39 \\
18 & P2-2   &   09:54:50   &   +69:06:56   &        5.8 & 6.6 & 117:       & $-$                     & $-$                     &   8.27 &  8.44 & 8.50 & 8.42  & 8.52 \\
19 & P2-3   &   09:54:54   &   +69:10:23   &        1.7 & 7.9 & $-$       & $-$                     & $-$                     &   8.60 &  8.37 & 8.60 & 8.56  & 8.31 \\
20 &        &   09:54:54   &   +69:10:21   &        1.4 & 7.8 & 2:         & $-$                     & $-$                     &   8.34 &  8.52 & 8.52 & 8.51  & 8.52 \\
21 &        &   09:54:54   &   +69:10:19   &        5.1 & 7.8 & $-$        & $-$                     & $-$                     &   8.51 &  8.40 & 8.52 & 8.49  & 8.41 \\
22 &        &   09:54:54   &   +69:10:17   &        1.8 & 7.8 & $-$        & $-$                     & $-$                     &   8.38 &  8.51 & 8.46 & 8.50  & 8.54 \\
23 &        &   09:54:54   &   +69:10:23   &        6.0 & 7.9 & 14:           & $-$                 & $-$                     &   8.24 &  8.42 & 8.37 & 8.33  & 8.57 \\
24 & P3-1   &   09:55:44   &   +69:07:19   &        1.4 & 5.4 & 18$\pm$8    & 10500$^{+900}_{-700}$  & 7.95$^{+0.11}_{-0.13}$   &   8.15 &  8.54 & 8.54 & 8.56  & 8.54 \\
25 &        &   09:55:45   &   +69:07:18   &        1.6 & 5.5 & 34:        & $-$                     & $-$                     &   8.10 &  8.50 & 8.53 & 8.53  & 8.55 \\
26 &        &   09:55:45   &   +69:07:18   &        2.3 & 5.5 & 2:         & $-$                     & $-$                     &   8.28 &  8.70 & 8.58 & 8.65  & 8.49 \\
27 & P3-2   &   09:55:36   &   +69:07:48   &        1.6 & 5.1 & 36:        & $-$                     & $-$                     &   8.60 &  8.55 & 8.47 & 8.47  & 8.57 \\
28 &        &   09:55:36   &   +69:07:47   &        2.6 & 5.1 & $-$        & $-$                     & $-$                     &   8.61 &  8.56 & 8.47 & 8.45  & 8.51 \\
29 &        &   09:55:35   &   +69:07:50   &        1.0 & 5.1 & $-$        & $-$                     & $-$                     &   8.73 &  8.63 & 8.49 & 8.49  & 8.53 \\
30 & P3-3   &   09:55:21   &   +69:08:40   &        3.9 & 5.4 & $-$        & $-$                     & $-$                     &   8.35 &  8.50 & 8.55 & 8.50  & 8.53 \\
31 &        &   09:55:20   &   +69:08:44   &        5.3 & 5.4 & $-$        & $-$                     & $-$                     &   8.28 &  8.62 & 8.58 & 8.61  & 8.53 \\
32 &        &   09:55:18   &   +69:08:48   &        4.4 & 5.5 & $-$        & $-$                     & $-$                     &   8.26 &  8.44 & 8.33 & 8.31  & 8.55 \\
33 &        &   09:55:17   &   +69:08:51   &        1.5 & 5.5 & $-$        & 9000$^{+900}_{-600}$    & 8.13$^{+0.14}_{-0.17}$  &   8.35 &  8.54 & 8.56 & 8.55  & 8.55 \\
34 &        &   09:55:17   &   +69:08:52   &        1.9 & 5.6 & 16:        & $-$                     & $-$                     &   8.52 &  8.53 & 8.50 & 8.47  & 8.52 \\
35 &        &   09:55:17   &   +69:08:55   &        4.1 & 5.6 & 16:        & 8200$^{+700}_{-600}$    & 8.46$^{+0.15}_{-0.16}$  &   8.57 &  8.53 & 8.49 & 8.39  & 8.49 \\
36 &        &   09:55:16   &   +69:08:59   &        2.0 & 5.6 & $-$        & 8400$^{+1000}_{-600}$   & 8.39$^{+0.15}_{-0.17}$  &   8.52 &  8.51 & 8.49 & 8.42  & 8.51 \\
37 &        &   09:55:15   &   +69:09:01   &        5.2 & 5.7 & $-$        & $-$                     & $-$                     &   8.32 &  8.64 & 8.59 & 8.61  & 8.53 \\
38 & P4-1   &   09:55:25   &   +69:08:19   &        7.2 & 5.1 & 13:        & $-$                     & $-$                     &   8.35 &  8.54 & 8.55 & 8.55  & 8.56 \\
39 &        &   09:55:26   &   +69:08:17   &        2.5 & 5.1 & 18$\pm$8         & $-$                     & $-$                     &   8.47 &  8.56 & 8.56 & 8.54  & 8.55 \\
40 & P4-2   &   09:55:19   &   +69:08:29   &        2.5 & 5.1 & 26$\pm$6    & 10000$^{+400}_{-300}$   & 8.11$\pm0.05$           &   8.53 &  8.52 & 8.48 & 8.34  & 8.46 \\
41 &        &   09:55:17   &   +69:08:31   &        3.0 & 5.2 & $-$        & $-$                     & $-$                     &   8.47 &  8.48 & 8.44 & 8.41  & 8.54 \\
42 &        &   09:55:14   &   +69:08:34   &        2.5 & 5.2 & 6:         & $-$                     & $-$                     &   8.45 &  8.57 & 8.56 & 8.55  & 8.55 \\
43 &        &   09:55:19   &   +69:08:29   &        1.5 & 5.1 & $-$        & $-$                     & $-$                     &   8.60 &  8.62 & 8.57 & 8.49  & 8.40 \\
44 &        &   09:55:22   &   +69:08:25   &        2.7 & 5.1 & $-$        & $-$                     & $-$                     &   8.16 &  8.39 & 8.37 & 8.47  & 8.69 \\
45 & P5-1   &   09:56:05   &   +69:03:44   &        3.2 & 5.4 & 3:         & $-$                     & $-$                     &   8.52 &  8.58 & 8.55 & 8.51  & 8.50 \\
46 &        &   09:56:05   &   +69:03:45   &        1.0 & 5.4 & $-$        & $-$                     & $-$                     &   8.31 &  8.58 & 8.57 & 8.58  & 8.55 \\
47 & P5-2   &   09:56:01   &   +69:04:00   &        2.6 & 4.8 & 16:        & $-$                     & $-$                     &   8.56 &  8.53 & 8.53 & 8.44  & 8.51 \\
48 &        &   09:55:60   &   +69:04:03   &        2.0 & 4.8 & $-$        & $-$                     & $-$                      &   8.39 &  8.56 & 8.57 & 8.57  & 8.57 \\
\hline
\end{tabular}
\label{Oxygen-abundances}
\end{minipage}
\end{table*}

We checked for the effect of the correction for stellar absorption on
the oxygen abundances derived for our observed \ion{H}{ii} regions. The values
of $12+\log(\mbox{O}/\mbox{H})$ change by $0\mbox{--}0.04$ dex in most of our regions
for the direct, ONS, C, O3N2, and N2 methods. The exceptions are region 24,
where the results of the direct method increase by 0.08 dex with the correction,
and region 6, the one with the largest correction, where the oxygen abundance
derived with the ONS method increases by 0.13 dex. The results of the
P method are more sensitive to this correction, with six regions showing increments
larger than 0.10 dex: regions 7, 21, and 44, where the oxygen abundance increases
by $\sim0.15$ dex, and regions 6, 14, and 26, with increments of 0.49, 0.28, and 0.24,
respectively.

We have calculated the galactocentric distances of the observed \ion{H}{ii}
regions assuming a planar geometry for M81, with a rotation angle of the major
axis of M81 of $157\degr$, a disc inclination of $59\degr$ \citep{KO:00a}, and a
distance of $3.63\pm0.34$ Mpc \citep{Free:01a}. Our 48 \ion{H}{ii} regions cover a
range of galactocentric distances of 4.8--9.0 kpc. In order to increase this
range, we selected from the literature other observations of \ion{H}{ii} regions
in M81. This also allows us to look for observational effects on the derived
abundances. The final sample is composed of 116 \ion{H}{ii} regions spanning a
range of galactocentric distances of 3--33 kpc, where 48 \ion{H}{ii} regions are
from this work and the remaining 68 from the works of \citet{Gar:87a},
\citet*{Bre:99a}, \citet{Stan:10a}, and \citet{Patt:12a}. We applied the same
procedures explained above to derive physical conditions and oxygen abundances for
the \ion{H}{ii} regions from the literature, using the line intensities reported
in the original papers. We also recalculated the galactocentric distances of these
\ion{H}{ii} regions using the same parameters stated above for M81. The results are
presented in Tables~\ref{ON1} and \ref{ON2}.

\begin{table*}
\begin{minipage}{180mm}
\caption{Oxygen and nitrogen abundances for the regions observed by
\citet{Patt:12a} and \citet{Stan:10a}.}
\begin{tabular}{lccccccccccc}
\hline
\multicolumn{1}{l}{ID} &
\multicolumn{1}{c}{$R$} &
\multicolumn{1}{c}{$T_{\rm e}$([\ion{N}{ii}])/$T_{\rm e}$([\ion{O}{iii}])} &
\multicolumn{6}{c}{$12+\log(\mbox{O}/\mbox{H})$}  & \multicolumn{3}{c}{$12+\log(\mbox{N}/\mbox{H})$}\\   
&  (kpc) &   (K)  & ($T_{\rm e}$) & (P) & (ONS) & (C) & (O3N2) & (N2) &  ($T_{\rm e}$) & (ONS) & (C) \\ 
\hline
\multicolumn{12}{c}{\citet{Patt:12a}}\\
02      &   22.6  &   $-$                                       &  $-$                     &  8.08 &  8.43 &  8.53 &  8.42 &  8.46 &  $-$		     & 7.33  & 7.40  \\
03      &   22.2  &   $-$                                       &  $-$                     &  8.25 &  8.43 &  8.40 &  8.37 &  8.46 &  $-$		     & 7.41  & 7.40  \\
07      &   22.8  &   $-$                                       &  $-$                     &  7.78 &  8.51 &  8.52 &  8.54 &  8.48 &  $-$		     & 7.32  & 7.24  \\
14      &   14.6  &   $-$                                       &  $-$                     &  8.41 &  8.57 &  8.44 &  8.29 &  8.39 &  $-$		     & 7.54  & 7.45  \\
17      &   21.6  &   $-$                                       &  $-$                     &  8.29 &  8.39 &  8.37 &  8.26 &  8.27 &  $-$		     & 7.07  & 7.03  \\
21      &   15.9  &   $14100\pm3800/11200^{+1000}_{-700}$       &  $8.16^{+0.13}_{-0.10}$  &  8.27 &  8.48 &  8.33 &  8.20 &  8.34 &  $7.41^{+0.17}_{-0.24}$ & 7.47  & 7.39  \\
24      &   16.1  &   $-$                                       &  $-$                     &  8.28 &  8.56 &  8.43 &  8.32 &  8.42 &  $-$		     & 7.45  & 7.36  \\
25      &   15.0  &   $-$                                       &  $-$                     &  8.07 &  8.49 &  8.45 &  8.50 &  8.52 &  $-$		     & 7.40  & 7.41  \\
26      &   31.1  &   $-$                                       &  $-$                     &  8.33 &  8.35 &  8.29 &  8.23 &  8.28 &  $-$		     & 7.11  & 7.05  \\
28      &   31.4  &   $-/12700^{+ 900}_{-700}$                  &  $8.19^{+0.06}_{-0.07}$  &  8.15 &  8.41 &  8.24 &  8.12 &  8.26 &  $7.23^{+0.07}_{-0.08}$ & 7.33  & 7.24  \\
29      &   29.2  &   $-$                                       &  $-$                     &  7.99 &  8.24 &  8.47 &  8.50 &  8.36 &  $-$		     & 6.93  & 7.08  \\
33      &   32.7  &   $-$                                       &  $-$                     &  8.23 &  8.41 &  8.21 &  8.24 &  8.27 &  $-$		     & 7.21  & 6.87  \\
35      &   24.1  &   $-$                                       &  $-$                     &  7.62 &  8.33 &  8.28 &  8.38 &  8.51 &  $-$		     & 7.19  & 7.17  \\
37      &   21.9  &   $-$                                       &  $-$                     &  8.01 &  8.40 &  8.43 &  8.37 &  8.44 &  $-$		     & 7.26  & 7.27  \\
disc1   &   6.4   &   $7500^{+900 }_{-500}/7300^{+ 700}_{-400}$ &  $8.74^{+0.16}_{-0.20}$  &  8.21 &  8.45 &  8.49 &  8.47 &  8.56 &  $7.82^{+0.18}_{-0.22}$ & 7.61  & 7.66  \\
disc2   &   11.5  &   $-$                                       &  $-$                     &  8.16 &  8.46 &  8.47 &  8.48 &  8.53 &  $-$		     & 7.55  & 7.55  \\
disc3   &   10.2  &   $8500^{+2000}_{-900}/9500^{+1000}_{-600}$ &  $8.55^{+0.20}_{-0.25}$  &  8.24 &  8.42 &  8.43 &  8.32 &  8.48 &  $7.56^{+0.24}_{-0.30}$ & 7.49  & 7.51  \\
disc4   &   7.9   &   $7800^{+900 }_{-600}/-$                   &  $8.67^{+0.16}_{-0.19}$  &  8.29 &  8.46 &  8.53 &  8.46 &  8.53 &  $7.78^{+0.18}_{-0.21}$ & 7.62  & 7.68  \\
disc5   &   5.7   &   $7900^{+1000}_{-600}/-$                   &  $8.60^{+0.17}_{-0.19}$  &  8.42 &  8.48 &  8.47 &  8.44 &  8.53 &  $7.80^{+0.19}_{-0.22}$ & 7.71  & 7.72  \\
disc6   &   5.0   &   $-$                                       &  $-$                     &  8.39 &  8.48 &  8.50 &  8.46 &  8.53 &  $-$		     & 7.70  & 7.73  \\
disc7   &   2.9   &   $-$                                       &  $-$                     &  8.27 &  8.55 &  8.57 &  8.59 &  8.61 &  $-$		     & 7.86  & 7.88  \\
\multicolumn{12}{c}{\citet{Stan:10a}}\\
HII4    &    9.3  &   $10800^{+9200 }_{-2300}/-$                &  $8.12^{+0.40}_{-0.41}$  &  8.31 &  8.44 &  8.46 &  8.33 &  8.47 &  $7.32^{+0.45}_{-0.52}$ & 7.50  & 7.53  \\
HII5    &    8.9  &   $11100\pm300/-$ 	                        &  $8.06\pm0.04$	   &  7.95 &  8.44 &  8.52 &  8.49 &  8.56 &  $7.29\pm0.04$ 	     & 7.45  & 7.52  \\
HII21   &    8.7  &   $-$			                &  $-$		           &  7.63 &  8.36 &  8.48 &  8.46 &  8.60 &  $-$		     & 7.37  & 7.45  \\
HII31   &    8.8  &   $8400^{+10300}_{-1400}/-$                 &  $8.59^{+0.40}_{-0.81}$  &  7.93 &  8.46 &  8.53 &  8.52 &  8.56 &  $7.63^{+0.54}_{-1.02}$ & 7.46  & 7.51  \\
HII42   &    9.0  &   $-$			                &  $-$		           &  7.88 &  8.41 &  8.45 &  8.47 &  8.57 &  $-$	             & 7.41  & 7.46  \\
HII72   &    6.9  &   $10300^{+3400}_{-1400}/-$                 &  $8.12^{+0.21}_{-0.26}$  &  8.32 &  8.45 &  8.47 &  8.34 &  8.43 &  $7.21^{+0.25}_{-0.30}$ & 7.42  & 7.43  \\
HII78   &    9.0  &   $-$			                &  $-$		           &  7.82 &  8.52 &  8.53 &  8.61 &  8.65 &  $-$	             & 7.58  & 7.62  \\
HII79   &    8.3  &   $9200^{+3200}_{-1200}/-$                  &  $8.25^{+0.27}_{-0.42}$  &  8.30 &  8.45 &  8.50 &  8.38 &  8.47 &  $7.32^{+0.31}_{-0.48}$ & 7.47  & 7.51  \\
HII81   &    7.2  &   $9100^{+10900}_{-1600}/-$                 &  $8.39^{+0.41}_{-0.86}$  &  8.07 &  8.41 &  8.48 &  8.42 &  8.52 &  $7.42^{+0.52}_{-1.11}$ & 7.42  & 7.47  \\
HII123  &    7.9  &   $8900^{+1000}_{-700}/-$                   &  $8.46^{+0.14}_{-0.17}$  &  8.14 &  8.43 &  8.54 &  8.42 &  8.48 &  $7.45^{+0.16}_{-0.19}$ & 7.40  & 7.47  \\
HII133  &    6.9  &   $11800^{+400}_{-300}/-$                   &  $7.95\pm0.04$	   &  8.22 &  8.42 &  8.47 &  8.38 &  8.50 &  $7.21\pm0.04$ 	     & 7.47  & 7.52  \\
HII201  &    6.9  &   $-/13300^{+2400}_{-1300}$                 &  $7.83^{+0.10}_{-0.13}$  &  8.48 &  8.51 &  8.36 &  8.28 &  8.40 &  $7.18^{+0.12}_{-0.14}$ & 7.60  & 7.50  \\
HII213  &    9.7  &   $-$			                &  $-$		           &  7.76 &  8.39 &  8.50 &  8.46 &  8.56 &  $-$		     & 7.35  & 7.42  \\
HII228  &   10.1  &   $9300\pm300/-$ 	                        &  $8.36\pm0.06$	   &  8.01 &  8.46 &  8.50 &  8.49 &  8.53 &  $7.44^{+0.06}_{-0.07}$ & 7.44  & 7.47  \\
HII233  &    5.9  &   $-$			                &  $-$		           &  8.02 &  8.48 &  8.53 &  8.53 &  8.58 &  $-$		     & 7.55  & 7.61  \\
HII249  &   10.6  &   $-$			                &  $-$		           &  7.49 &  8.34 &  8.48 &  8.44 &  8.57 &  $-$		     & 7.27  & 7.34  \\
HII262  &    9.9  &   $11500^{+1100}_{-800}/-$                  &  $8.19^{+0.11}_{-0.13}$  &  7.69 &  8.36 &  8.45 &  8.44 &  8.57 &  $7.31^{+0.12}_{-0.14}$ & 7.32  & 7.40  \\
HII282  &    5.1  &   $-$			                &  $-$		           &  8.09 &  8.48 &  8.52 &  8.52 &  8.55 &  $-$		     & 7.54  & 7.57  \\
HII325  &    9.5  &   $10800\pm2500/-$	                        &  $8.14^{+0.46}_{-0.23}$  &  8.15 &  8.41 &  8.45 &  8.37 &  8.47 &  $7.22^{+0.47}_{-0.36}$ & 7.36  & 7.40  \\
HII352  &   10.7  &   $-$			                &  $-$		           &  7.24 &  8.26 &  8.37 &  8.38 &  8.61 &  $-$		     & 7.25  & 7.32  \\
HII384  &    7.0  &   $-$			                &  $-$		           &  8.14 &  8.49 &  8.51 &  8.52 &  8.54 &  $-$		     & 7.56  & 7.59  \\
HII403  &    9.9  &   $9400^{+2400}_{-1100}/-$                  &  $8.57^{+0.23}_{-0.31}$  &  7.74 &  8.35 &  8.44 &  8.41 &  8.54 &  $7.49^{+0.26}_{-0.35}$ & 7.27  & 7.34  \\
\hline
\end{tabular}
\label{ON1}
\end{minipage}
\end{table*}

\begin{table*}
\caption{Oxygen and nitrogen abundances for the regions observed by \citet{Bre:99a} and
\citet{Gar:87a}.}
\begin{tabular}{lcccccccc}
\hline
\multicolumn{1}{l}{ID} &
\multicolumn{1}{c}{$R$} &
 \multicolumn{5}{c}{$12+\log(\mbox{O}/\mbox{H})$}  & \multicolumn{2}{c}{$12+\log(\mbox{N}/\mbox{H})$} \\
 &  (kpc)   & (P) & (ONS) & (C) & (O3N2) & (N2) & (ONS) & (C)  \\ 
\hline
\multicolumn{9}{c}{\citet{Bre:99a}}\\
GS1       & 5.5        & 8.35 & 8.48 & 8.53 &  8.45 & 8.50  & 7.60 &  7.64   \\
GS2       & 4.8        & 8.35 & 8.47 & 8.51 &  8.43 & 8.50  & 7.57 &  7.59     \\
GS4       & 8.6        & 8.24 & 8.43 & 8.49 &  8.34 & 8.46  & 7.41 &  7.45   \\
GS7       & 9.0        & 8.18 & 8.44 & 8.49 &  8.46 & 8.54  & 7.52 &  7.56   \\
GS9       & 6.5        & 8.40 & 8.49 & 8.50 &  8.49 & 8.57  & 7.77 &  7.79   \\
GS11      & 5.6        & 8.51 & 8.50 & 8.49 &  8.41 & 8.52  & 7.75 &  7.75   \\
GS12      & 5.0        & 8.14 & 8.47 & 8.54 &  8.49 & 8.52  & 7.50 &  7.54   \\
GS13      & 4.8        & 8.52 & 8.57 & 8.58 &  8.54 & 8.54  & 7.92 &  7.93   \\
M\"unch1  & 16.0       & 8.13 & 8.47 & 8.29 &  8.16 & 8.33  & 7.43 &  7.34   \\
M\"unch18 & 10.1       & 8.48 & 8.56 & 8.37 &  8.28 & 8.42  & 7.73 &  7.62   \\
\multicolumn{9}{c}{\citet{Gar:87a}}\\
HK105    &     9.2   & 7.99 & 8.48 & 8.44 & 8.50 &  8.50  & 7.39 & 7.36 \\
HK152    &     5.6   & 8.41 & 8.51 & 8.49 & 8.46 &  8.48  & 7.63 & 7.62 \\
HK230    &     4.8   & 8.58 & 8.57 & 8.47 & 8.51 &  8.55  & 7.98 & 7.94  \\
HK268    &     5.5   & 8.48 & 8.51 & 8.53 & 8.45 &  8.51  & 7.72 & 7.75 \\
HK305-12 &     5.1   & 8.48 & 8.50 & 8.49 & 8.41 &  8.51  & 7.73 & 7.73 \\
HK343-50 &     4.8   & 8.40 & 8.48 & 8.47 & 8.43 &  8.50  & 7.63 & 7.63 \\
HK453    &     5.0   & 8.21 & 8.48 & 8.47 & 8.49 &  8.52  & 7.54 & 7.54 \\
HK472    &     4.0   & 8.39 & 8.53 & 8.47 & 8.56 &  8.60  & 7.94 & 7.88 \\
HK500    &     5.6   & 8.57 & 8.54 & 8.39 & 8.35 &  8.49  & 7.82 & 7.78 \\
HK652    &     6.5   & 8.47 & 8.51 & 8.45 & 8.48 &  8.56  & 7.82 & 7.81 \\
HK666    &     7.0   & 8.37 & 8.46 & 8.42 & 8.37 &  8.50  & 7.62 & 7.59 \\
HK712    &     7.0   & 8.54 & 8.50 & 8.38 & 8.31 &  8.42  & 7.64 & 7.57 \\
HK741    &     9.0   & 8.29 & 8.47 & 8.49 & 8.45 &  8.53  & 7.54 & 7.56 \\
HK767    &     8.6   & 8.30 & 8.44 & 8.49 & 8.35 &  8.48  & 7.47 & 7.54 \\
M\"unch18  &   10.1   & 8.47 & 8.51 & 8.36 & 8.31 &  8.50  & 7.85 & 7.78 \\
\hline
\end{tabular}
\label{ON2}
\end{table*}

\begin{figure*}
\includegraphics[width=0.85\textwidth, trim=20 0 15 0, clip=yes]{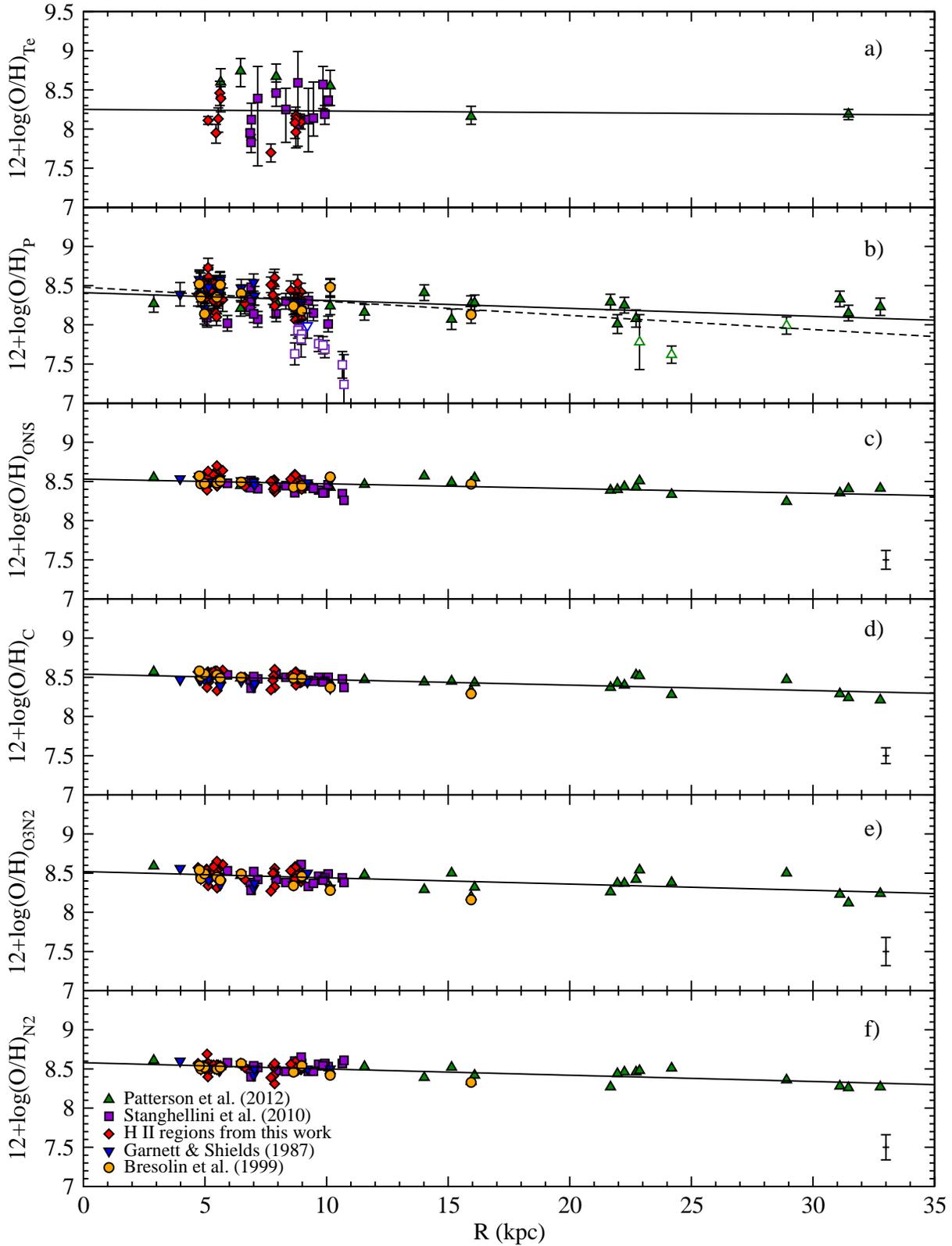}
\caption{Oxygen abundances in \ion{H}{ii} regions of M81 as a function of their
galactocentric distances and the abundance gradients resulting from our fits.
Panels (a) to (f) show the results of the direct method and the methods P, ONS, C,
O3N2, and N2. The different symbols indicate the references for the observational
data we used, and are identified in panel (f). Panels (c) to (f) show in the lower
right corner the typical uncertainty in the oxygen abundances derived with the
corresponding method. In panel (b) we also plot with a discontinuous line the
gradient fitted when the regions where the P method is not working (plotted as
empty symbols; see text) are included in the fit. Note that all the panels
are at the same scale.
}
\label{gradients}
\end{figure*}

The direct method could be applied to 31 \ion{H}{ii} regions of the final sample
where the electron temperature can be estimated ($T_{\rm e}$([\ion{N}{ii}]),
$T_{\rm e}$([\ion{O}{iii}]), or both): 12 from this work, 13 from \citet{Stan:10a}
and six from \citet{Patt:12a}. The strong-line methods were applied to all the
regions in the final sample.
Fig.~\ref{gradients} shows the oxygen abundances obtained with the different
methods we are using as a function of galactocentric distance for the \ion{H}{ii}
regions in our final sample. Panel (a) shows the results for the 31 \ion{H}{ii}
regions with some temperature estimate that allows us to use the direct method;
panels (b) to (f) show the results obtained with the strong-line methods for the 116
\ion{H}{ii} regions of the whole sample. In panel (b) we plot with open symbols
the results for the 14 regions that are classified as belonging to the upper branch
of the metallicity relation, but whose values of $12+\log(\mbox{O}/\mbox{H})$, derived
with this relation, fall below 8.0, the region of the lower branch.

We fitted straight lines with the least-squares method to the data in
Fig.~\ref{gradients} in order to derive the abundance gradient implied by each of
the methods used for the abundance determination. Weighted least-squares fits
produce similar values for the parameters, but we present the non-weighted results
because some of the data seem to be affected by systematic errors, and we do not
think that a robust estimation is required for our purposes. The fits are plotted
in Fig.~\ref{gradients}, and the parameters of the fitted gradients are listed in
Table~\ref{Results}, where we list for each method the number of regions used
($N$), the intercept and the slope of the fit, and the standard deviation of the
points from this fit.
In the case of the P method, we excluded from the fit the regions where
this method does not seem to be working properly (see above). The discontinuous line
in panel (b) shows the results when these regions are included. The intercept and slope
for this fit are $8.48\pm0.03$ and $-0.018\pm0.004$, respectively, with a dispersion of
0.24 dex.

\begin{table}
\caption{Oxygen abundance gradients and dispersions.}
\begin{tabular}{lcccc}
\hline
\multicolumn{1}{l}{Method} & \multicolumn{1}{c}{$N$} &
\multicolumn{1}{c}{12+log(O/H)$_0$} &
\multicolumn{1}{c}{$\frac{\Delta(\log({\rmn O}/{\rmn H}))}{\Delta(R)}$} &
\multicolumn{1}{c}{$\sigma$} \\
\multicolumn{3}{l}{} &
\multicolumn{1}{c}{(dex kpc$^{-1}$)} & \multicolumn{1}{l}{} \\
\hline
$T_{\rm e}$ &\ 31 & $8.26\pm0.10$ &$-0.002\pm0.010$  & 0.25 \\
P           & 102 & $8.41\pm0.03$ &$-0.010\pm0.003$  & 0.15 \\
ONS         & 116 & $8.53\pm0.01$ &$-0.006\pm0.001$  & 0.07 \\
C           & 116 & $8.54\pm0.01$ &$-0.007\pm0.001$  & 0.06 \\
O3N2        & 116 & $8.52\pm0.02$ &$-0.008\pm0.001$  & 0.09 \\
N2          & 116 & $8.58\pm0.01$ &$-0.008\pm0.001$  & 0.06 \\
\hline
\end{tabular}
\label{Results}
\end{table}

\subsection{Nitrogen abundances and the N/O abundance gradient}

The N/H and N/O abundance ratios were calculated using the direct method for
31 \ion{H}{ii} regions and the ONS and C methods for the whole sample.
Tables~\ref{ON1}, \ref{ON2}, and \ref{nitrogen-abundances} show the results.

\begin{table*}
\caption{The nitrogen abundances derived with the direct method ($T_{\rm e}$) and two strong-line methods
(ONS and C) for the 48 regions in our observed sample.}
\begin{tabular}{lcccccc}
\hline
\multicolumn{1}{c}{ID} & \multicolumn{3}{c}{$12+\log(\mbox{N}/\mbox{H})$}
& \multicolumn{3}{c}{$\log(\mbox{N}/\mbox{O})$} \\
& ($T_{\rm e}$) & (ONS) & (C) & ($T_{\rm e}$) & (ONS) & (C) \\
\hline
1   &  $7.39^{+0.08}_{-0.09}$  &  7.58  &  7.58  &  $-0.76\pm0.05$           &  $-0.87$ & $-0.85$ \\
2   &  $7.35^{+0.10}_{-0.12}$  &  7.53  &  7.59  &  $-0.75\pm0.06$           &  $-0.91$ & $-0.93$ \\
3   &  $-$ 		       &  7.61  &  7.62  &  $-$			     &  $-0.87$ & $-0.87$ \\
4   &  $-$ 		       &  7.46  &  7.50  &  $-$			     &  $-0.94$ & $-0.94$ \\
5   &  $7.34^{+0.18}_{-0.23}$  &  7.62  &  7.69  &  $-0.63^{+0.10}_{-0.09}$  &  $-0.84$ & $-0.86$ \\
6   &  $-$ 		       &  7.70  &  7.65  &  $-$			     &  $-0.83$ & $-0.82$ \\
7   &  $-$ 		       &  7.54  &  7.56  &  $-$			     &  $-0.92$ & $-0.93$ \\
8   &  $7.50^{+0.13}_{-0.14}$  &  7.68  &  7.68  &  $-0.68^{+0.07}_{-0.06}$  &  $-0.81$ & $-0.78$ \\
9   &  $-$ 		       &  7.73  &  7.72  &  $-$			     &  $-0.84$ & $-0.85$ \\
10  &  $7.48^{+0.09}_{-0.10}$  &  7.65  &  7.64  &  $-0.65\pm0.06$           &  $-0.79$ & $-0.76$ \\
11  &  $-$ 		       &  7.70  &  7.72  &  $-$			     &  $-0.80$ & $-0.78$ \\
12  &  $7.34^{+0.19}_{-0.24}$  &  7.53  &  7.59  &  $-0.74^{+0.11}_{-0.09}$  &  $-0.90$ & $-0.92$ \\
13  &  $-$ 		       &  7.65  &  7.64  &  $-$			     &  $-0.94$ & $-0.92$ \\
14  &  $-$ 		       &  7.61  &  7.63  &  $-$			     &  $-0.90$ & $-0.90$ \\
15  &  $-$ 		       &  7.68  &  7.67  &  $-$			     &  $-0.91$ & $-0.89$ \\
16  &  $-$ 		       &  7.85  &  7.85  &  $-$			     &  $-0.68$ & $-0.67$ \\
17  &  $7.14^{+0.13}_{-0.15}$  &  7.65  &  7.56  &  $-0.56\pm0.08$	     &  $-0.85$ & $-0.78$ \\ 
18  &  $-$ 		       &  7.55  &  7.59  &  $-$			     &  $-0.89$ & $-0.91$ \\
19  &  $-$ 		       &  7.58  &  7.71  &  $-$			     &  $-0.79$ & $-0.89$ \\ 
20  &  $-$ 		       &  7.66  &  7.67  &  $-$			     &  $-0.86$ & $-0.85$ \\ 
21  &  $-$ 		       &  7.48  &  7.62  &  $-$			     &  $-0.91$ & $-0.90$ \\ 
22  &  $-$ 		       &  7.73  &  7.71  &  $-$			     &  $-0.78$ & $-0.75$ \\ 
23  &  $-$ 		       &  7.71  &  7.70  &  $-$			     &  $-0.71$ & $-0.67$ \\ 
24  &  $7.35^{+0.14}_{-0.17}$  &  7.60  &  7.61  &  $-0.68\pm0.07$	     &  $-0.98$ & $-0.93$ \\
25  &  $-$ 		       &  7.52  &  7.53  &  $-$			     &  $-0.98$ & $-1.00$ \\
26  &  $-$ 		       &  7.74  &  7.64  &  $-$			     &  $-0.96$ & $-0.94$ \\
27  &  $-$ 		       &  7.99  &  7.95  &  $-$			     &  $-0.56$ & $-0.52$ \\
28  &  $-$ 		       &  7.89  &  7.84  &  $-$			     &  $-0.68$ & $-0.63$ \\
29  &  $-$ 		       &  8.16  &  8.07  &  $-$			     &  $-0.48$ & $-0.42$ \\
30  &  $-$ 		       &  7.68  &  7.72  &  $-$			     &  $-0.82$ & $-0.83$ \\
31  &  $-$ 		       &  7.76  &  7.73  &  $-$			     &  $-0.86$ & $-0.85$ \\
32  &  $-$ 		       &  7.74  &  7.73  &  $-$			     &  $-0.70$ & $-0.60$ \\ 
33  &  $7.52^{+0.16}_{-0.21}$  &  7.76  &  7.77  &  $-0.62\pm0.09$	     &  $-0.79$ & $-0.79$ \\ 
34  &  $-$ 		       &  7.80  &  7.79  &  $-$			     &  $-0.73$ & $-0.71$ \\
35  &  $7.73^{+0.17}_{-0.19}$  &  7.79  &  7.75  &  $-0.75^{+0.09}_{-0.08}$  &  $-0.74$ & $-0.74$ \\ 
36  &  $7.69^{+0.18}_{-0.21}$  &  7.76  &  7.75  &  $-0.71\pm0.10$	     &  $-0.75$ & $-0.74$ \\ 
37  &  $-$ 		       &  7.81  &  7.78  &  $-$			     &  $-0.83$ & $-0.81$ \\ 
38  &  $-$ 		       &  7.79  &  7.81  &  $-$			     &  $-0.75$ & $-0.74$ \\
39  &  $-$ 		       &  7.86  &  7.87  &  $-$			     &  $-0.69$ & $-0.69$ \\
40  &  $7.47^{+0.06}_{-0.07}$  &  7.71  &  7.69  &  $-0.65\pm0.04$	     &  $-0.81$ & $-0.79$ \\
41  &  $-$ 		       &  7.77  &  7.76  &  $-$			     &  $-0.72$ & $-0.68$ \\
42  &  $-$ 		       &  7.88  &  7.88  &  $-$			     &  $-0.70$ & $-0.68$ \\
43  &  $-$ 		       &  7.89  &  7.74  &  $-$			     &  $-0.73$ & $-0.83$ \\
44  &  $-$ 		       &  7.78  &  7.82  &  $-$			     &  $-0.60$ & $-0.55$ \\
45  &  $-$ 		       &  7.83  &  7.81  &  $-$			     &  $-0.75$ & $-0.74$ \\
46  &  $-$ 		       &  7.78  &  7.78  &  $-$			     &  $-0.80$ & $-0.79$ \\
47  &  $-$ 		       &  7.79  &  7.81  &  $-$			     &  $-0.74$ & $-0.72$ \\
48  &  $-$ 		       &  7.85  &  7.87  &  $-$			     &  $-0.71$ & $-0.70$ \\
\hline
\end{tabular}
\label{nitrogen-abundances}
\end{table*}

\begin{figure*}
\includegraphics[width=0.90\textwidth, trim=10 0 10 0, clip=yes]{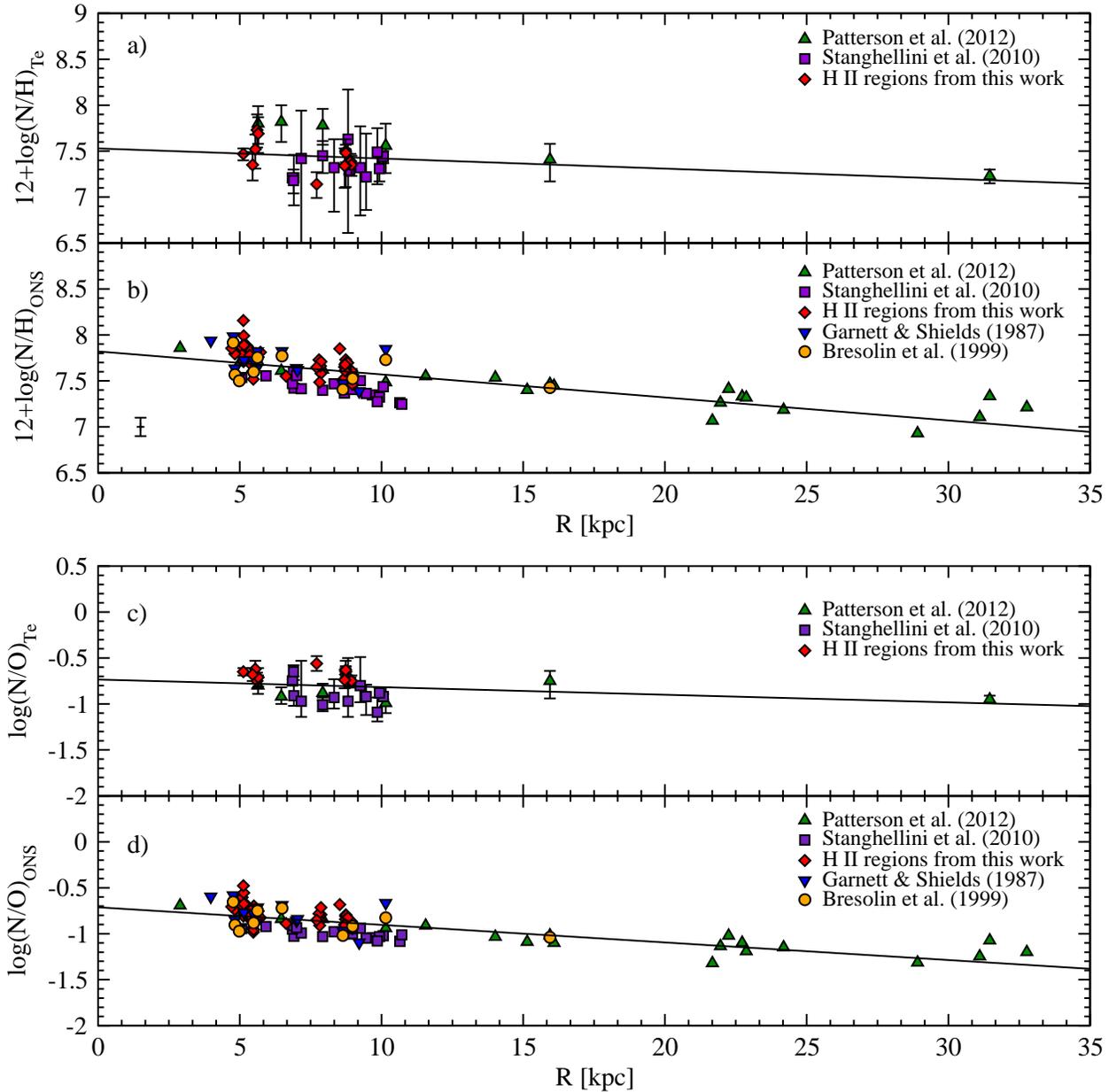}
\caption{N/H and N/O abundances in the \ion{H}{ii} regions of M81 as a function of their
galactocentric distances and the abundance gradients resulting from the fits. Panels~(a) and
(c) show the results of the direct method, and panels~(b) and (d) the results of the
ONS method. The different symbols indicate the references for the observational data we used. In
all panels, the vertical scale spans the same range in orders of magnitude displayed in
Fig.~\ref{gradients}.} 
\label{N-gradient}
\end{figure*}

Fig.~\ref{N-gradient} shows the results for the N/H and N/O abundances as a function of
galactocentric distance. Panels~(a) and (c) are for the abundances obtained with the direct
method and panels~(b) and (d) those for the ONS method. For ease of comparison, the panels
cover the same range in orders of magnitude that we used in Fig.~\ref{gradients}.
We have not plotted the results of the
C method, because they show a similar distribution of values to those of the ONS method. The
least-squares fits to the data are also plotted in the figure, and in
Table~\ref{Results-nitrogen} we list for each method the number of regions used in the fits,
the derived intercepts and slopes, and the dispersions around the gradients. The slopes
obtained with the ONS and C methods are very similar, $\sim-0.020$ dex kpc$^{-1}$, whereas
the direct method implies a shallower slope, $-0.008$ dex kpc$^{-1}$. The N/H abundance ratios
derived with the ONS and C methods can be assigned uncertainties of $\sim0.10\mbox{--}0.15$
dex. The methods do not provide estimates of the uncertainties in the derived N/O abundance
ratios, but the dispersions around the gradients implied by these methods suggest that the
random uncertainties are $\sim0.1$~dex.

\begin{table*}
\caption{N/H and N/O abundance gradients and dispersions}
\begin{tabular}{lccccccc}
\hline
\multicolumn{1}{l}{Method} & \multicolumn{1}{c}{$N$} &
\multicolumn{1}{c}{12+log(N/H)$_0$} &
\multicolumn{1}{c}{$\frac{\Delta(\log({\rmn N}/{\rmn H}))}{\Delta(R)}$} &
\multicolumn{1}{c}{$\sigma$} &
\multicolumn{1}{c}{log(N/O)$_0$} &
\multicolumn{1}{c}{$\frac{\Delta(\log({\rmn N}/{\rmn O}))}{\Delta(R)}$} &
\multicolumn{1}{c}{$\sigma$} \\
\multicolumn{3}{l}{} &
\multicolumn{1}{c}{(dex kpc$^{-1}$)} & \multicolumn{1}{l}{} &
\multicolumn{1}{l}{} &
\multicolumn{1}{c}{(dex kpc$^{-1}$)} & \multicolumn{1}{l}{}\\
\hline
$T_{\rm e}$ & 31 & 7.53$\pm$0.07  & $-$0.011$\pm$0.007 & 0.18 &  $-$0.73$\pm$0.05  & $-$0.008$\pm$0.005 & 0.13 \\
ONS         & 116 & 7.82$\pm$0.03  & $-$0.025$\pm$0.002 & 0.15 &  $-$0.71$\pm$0.05  & $-$0.019$\pm$0.002 & 0.11 \\
C           & 116 & 7.85$\pm$0.02  & $-$0.020$\pm$0.002 & 0.12 &  $-$0.69$\pm$0.05  & $-$0.020$\pm$0.002 & 0.13  \\
\hline
\end{tabular}
\label{Results-nitrogen}
\end{table*}

\subsection{Comparison with other works}

The values that we obtain for the slope of the metallicity gradient go from 
$-0.010$ to $-0.002$ dex kpc$^{-1}$, smaller in absolute values than most other
determinations of the oxygen abundance gradient in M81. Table~\ref{Literature}
provides a compilation of some previous results ordered chronologically, where we
list the method and number of regions used in each case, the range of
galactocentric distances covered by the objects and the intercept and the slope of
the fits. Besides two old determinations based on the R$_{23}$ method calibrated
with photoionization models by \citet{Pa:79a}, we have chosen to present the
results that are based on methods similar to the ones we use. The most recent
determination, that of \citet{Pil:14a}, is based on abundances calculated with the
P and C methods slightly modified, which we label as P$^\prime$ and C$^\prime$.
\citet{Pil:14a} also derived the gradient for N/H with their C$^\prime$ method
for regions with galactocentric distances in the range 4--13 kpc, finding a slope of $-0.033$,
steeper than the one we find with the C method for the range of 3--33 kpc, $-0.020$.

\begin{table}
\caption{Oxygen abundance gradients from the literature.}
\begin{tabular}{lclllc}
\hline
\multicolumn{1}{l}{Method} & \multicolumn{1}{c}{$N$} &
\multicolumn{1}{c}{$\Delta R$} &
\multicolumn{1}{c}{log(O/H)$_0$} &
\multicolumn{1}{c}{$\frac{\Delta(\log({\rmn O}/{\rmn H}))}{\Delta(R)}$} & Ref. \\
\multicolumn{2}{l}{} & \multicolumn{1}{l}{(kpc)} &
\multicolumn{1}{l}{+12} &
\multicolumn{1}{c}{(dex kpc$^{-1}$)} & \multicolumn{1}{l}{} \\
\hline
R$_{23}$              & 10  & \ 4--8  & \ --          & $-$0.045           & 1 \\
R$_{23}$              & 18  & \ 3--15 & \ --          & $-$0.08            & 2 \\
P                     & 36  & \ 4--12 & 8.69          &  $-$0.031 & 3 \\
$T_{\rm e}$           & 31  & \ 4--17 & 9.37$\pm$0.24 & $-$0.093$\pm$0.020 & 4 \\
P                     & 21  & \ 3--33 & 8.34$\pm$0.12 & $-$0.013$\pm$0.006 & 5 \\
P                     & 49  & \ 3--33 & 8.47$\pm$0.06 & $-$0.016$\pm$0.006 & 5 \\
$T_{\rm e}$           &\ 7  & \ 6--32 & 8.76$\pm$0.13 & $-$0.020$\pm$0.006 & 5 \\
$T_{\rm e}$           & 28  & \ 5--10 & 9.20$\pm$0.11 & $-$0.088$\pm$0.013 & 6 \\
P$^\prime$+C$^\prime$ & --  & \ 4--13 & 8.58$\pm$0.02 & $-$0.011$\pm$0.003 & 7 \\
\hline
\end{tabular}
\label{Literature}
References: (1) \citet{Stau:84a}, (2) \citet{Gar:87a}, (3) \citet{Pil:04a}, (4)
\citet{Stan:10a}, (5) \citet{Patt:12a}, (6) \citet{Stan:14a}, (7) \citet{Pil:14a}.
\end{table}

The results shown in Figs.~\ref{gradients} and \ref{N-gradient}, and
Tables~\ref{Results}, \ref{Results-nitrogen}, and \ref{Literature} illustrate the well-known
fact that gradient determinations are very sensitive to the method, to the number of objects
used, and to the range of galactocentric distances covered by these objects.

Our results with the P method
are very similar to those obtained by \citet{Patt:12a} with this method, and this
is the case where both the procedure followed in the abundance determination and
the range of galactocentric distances covered agree more closely. \citet{Patt:12a}
use larger error bars than we do for the results of the P method and do a weighted
least-squares fit, but the main difference between their results and ours is in the
abundances obtained for the \ion{H}{ii} regions observed by \citet{Stan:10a}, which
they also use. The oxygen abundances that we derive for these regions are lower
than the ones they find. This is clearly seen in panel (b) of Fig.~\ref{gradients}
where several objects located between 8 and 11 kpc have oxygen abundances much
lower than $12+\log(\mbox{O}/\mbox{H})=8.0$, whereas \citet{Patt:12a} find
$12+\log(\mbox{O}/\mbox{H})>8.0$ for all these regions. These differences are
partly due to the fact that \citet{Patt:12a} do not include all the \ion{H}{ii}
regions of \citet{Stan:10a}, but use only those for which there is also an estimate
of the electron temperature. However, we can only reproduce their results for the
\ion{H}{ii} regions in common if we use the line ratios of \citet{Stan:10a}
uncorrected for extinction, which \citet{Patt:12a} seem to have inadvertently done.
The results we derive with the P method for the \ion{H}{ii} regions
observed by \citet{Patt:12a} agree within 0.01 dex with the ones derived by these
authors with the exception of four objects which belong to the upper branch of the
metallicity calibration according to our classification scheme (see Section 3.2.1),
but are in an ambiguous region according to the procedure followed by
\citet{Patt:12a}. For these regions they calculate an average of the oxygen
abundances implied by the upper and lower branch of the calibration, obtaining
values that differ from the ones we calculated, using their line intensities,
by 0.05--0.26 dex.

On the other hand, there are several \ion{H}{ii} regions in our full sample which
are classified as belonging to the upper branch following both our classification
scheme and the one used by \citet{Patt:12a}, but whose abundances, calculated using
the upper-branch relation of the P method, lie in the region that should be covered
by the lower branch [all the regions with $12+\log(\mbox{O}/\mbox{H})\le8.0$ in
Fig.~\ref{gradients}(b), which are plotted as empty symbols, and in the lower panel
of fig. 10 of \citealt{Patt:12a}].
Our observed regions do not show this problem, but two of them, regions 14
and 44, would have the same behaviour if we had not corrected their spectra for the
effects of stellar absorption: the uncorrected spectra imply values of
$12+\log(\mbox{O}/\mbox{H})=7.76$ and 8.00, whereas the corrected spectra change
those values to 8.06 and 8.16, respectively. Since neither \citet{Patt:12a} nor
\citet{Stan:10a} correct their spectra for stellar absorption, the regions they
observed where the P method has problems might also be affected in the same way.
We do not consider in our fit of Table~\ref{Results} the \ion{H}{ii} regions where
the P method is not working properly. If we include them, we get an intercept and
slope for the gradient of $8.48\pm0.03$ and $-0.018\pm0.004$, respectively, with a
dispersion of 0.24 dex. This fit is plotted with a discontinuous line in
Fig.~\ref{gradients}(b). \citealt{Patt:12a} did not reject from their fits the objects
that had problems with the P method, and the gradients they derive with this method
are intermediate between our two fits.

Our results for the abundances implied by the direct method in the \ion{H}{ii}
regions observed by \citet{Stan:10a} are significantly different from those derived
by these authors: we get oxygen abundances that are lower by up to 0.3 dex. The
differences are mainly due to the fact that \citet{Stan:10a} calculated the neutral
oxygen abundance in several objects using [\ion{O}{i}] emission and added it to the
O$^+$ and O$^{++}$ abundances to get the total oxygen abundance as can be seen
in their table~3, available online; see, for example, the results for their region
number 5. This is not a procedure usually followed for \ion{H}{ii} regions since the
ionization potentials of O$^0$ and H$^0$ are both $\simeq13.6$ eV, suggesting that
[\ion{O}{i}] emission should arise in regions close to the ionization front.
Besides, charge-exchange reactions between O$^0$ and H$^+$ tend to keep O$^0$ outside
the ionized region \citep{Os:06a}. The O$^0$/H$^+$ abundance ratios derived by
\citet{Stan:10a} are also very high, 30 to 230 times larger than the ones we estimate.
Since \citet{Patt:12a} compared their results with the direct method with those reported
by \citet{Stan:10a}, they found a better agreement of the two sets than the one
that can be observed in Fig.~\ref{gradients}. 

The differences between the abundances we derive with the direct method using
the line intensities of \citet{Patt:12a} and the values given by these authors are
below 0.2 dex, and seem to be due to typos in their tables. For example, \citet{Patt:12a}
give a value for $T_{\rm e}$([\ion{O}{iii}]) for their region 26, but no intensity is
provided for the [\ion{O}{iii}]~$\lambda4363$ line for this region in their table
2. In addition, some of the values they list for the total oxygen abundance in
their table~4, and plot in their figures, are transposed, namely the values of O/H
given for their regions disc1, disc3, and disc4. If we add the values of
O$^+$/H$^+$ and O$^{++}$/H$^+$ listed in their table~4 for each of these regions,
we get the total oxygen abundance that they attribute to a different region; for example,
the oxygen abundance implied by their ionic abundances in disc3 is assigned by
them to region disc4.
These differences, along with the fact that our observations
lead to lower oxygen abundances for the galactocentric range in common with the
other samples, explain the very different value that we obtain with the direct
method for the abundance gradient, $-0.002$ dex kpc$^{-1}$ versus $-0.020$ dex
kpc$^{-1}$ \citep[the result of][that covers a range of galactocentric distances
similar to ours]{Patt:12a}. An inspection of Fig.~\ref{gradients}(a) shows
that the inclusion of data from different works is the main reason of this difference:
we would get a steeper gradient if we only used the data obtained by \citet{Patt:12a}.

We have several regions in common with other authors, and Table~\ref{comparison}
shows a comparison between the oxygen and nitrogen abundances we derive with different methods using
the line intensities reported for each region. The apertures are different, and in
two cases we extracted the spectra of two knots at the positions covered by other
works, but the differences in the abundances implied by each method are of the same
order as the differences that we find in Figs.~\ref{gradients} and \ref{N-gradient} for regions at
similar galactocentric distances. Since these differences depend on the method and
in some cases are larger than the estimated uncertainties, we think that
the results in  Table~\ref{comparison} and Figs.~\ref{gradients} and \ref{N-gradient} illustrate the
robustness of the methods to different observational problems that are not
necessarily included in the estimates of the uncertainties in the line intensities.
The data obtained by different authors will be affected in different amounts by
uncertainties that are difficult to estimate, such as those introduced by atmospheric
differential refraction \citep{Fil:82}, flux calibration or extraction, extinction
correction, and the measurement of weak lines in spectra that are not deep enough or
have poor spectral resolution. Those methods that give consistent results when applied to
different sets of observations can be considered more robust to these observational
effects.

\begin{table*}
\begin{minipage}{150mm}
\caption{Comparison of our results for the \ion{H}{ii} regions in common with other
samples.}
\begin{tabular}{lccccccccccc}
\hline
\multicolumn{1}{l}{ID} & \multicolumn{1}{c}{Ref.} &
\multicolumn{1}{c}{$T_{\rm e}$([\ion{N}{ii}])} &
\multicolumn{6}{c}{$12+\log(\mbox{O}/\mbox{H})$} &
\multicolumn{3}{c}{$12+\log(\mbox{N}/\mbox{H})$}  \\

 & & (K) & ($T_{\rm e}$) & (P) & (ONS) & (C) & (O3N2) & (N2) & ($T_{\rm e}$) & (ONS) & (C)  \\
\hline
1        & 1 & 10100$^{+500}_{-400}$   & $8.13^{+0.06}_{-0.07}$   & 8.33 & 8.45 & 8.43 & 8.42 & 8.53   & $7.39^{+0.08}_{-0.09}$  & 7.58  &  7.58 \\
HII31    & 2 & 8400$^{+10300}_{-1400}$ & $8.59^{+0.40}_{-0.81}$   & 7.93 & 8.46 & 8.53 & 8.52 & 8.56   & $7.63^{+0.54}_{-1.02}$  & 7.46  &  7.51 \\                         
GS7      & 3 & $-$                     & $-$                      & 8.18 & 8.44 & 8.49 & 8.46 & 8.54   & $-$  & 7.52 & 7.56 \\                           
HK741    & 4 & $-$                     & $-$                      & 8.29 & 8.47 & 8.49 & 8.45 & 8.53   & $-$  & 7.54 & 7.56  \\                          
\hline                                                                                                                            
15       & 1 & $-$                     & $-$                      & 8.28 & 8.58 & 8.56 & 8.57 & 8.52   &  $-$ 		         &  7.68  &  7.67 \\                           
GS4      & 3 & $-$ 		       & $-$		          & 8.24 & 8.43 & 8.49 & 8.34 & 8.46   &  $-$  & 7.41 &  7.45 \\                          
HK767    & 4 & $-$ 		       & $-$		          & 8.30 & 8.44 & 8.49 & 8.35 & 8.48   &  $-$  & 7.47 & 7.54  \\                          
\hline                                                                                                                            
35       & 1 &  8200$^{+700}_{-600}$   & 8.46$^{+0.15}_{-0.16}$   & 8.57 & 8.53 & 8.49 & 8.39 & 8.49   &  $7.73^{+0.17}_{-0.19}$  &  7.79  & 7.75  \\                        
disc5    & 5 & 7900$^{+1000}_{-600}$   & 8.60$^{+0.17}_{-0.19}$   & 8.42 & 8.48 & 8.47 & 8.44 & 8.53   &  $7.80^{+0.19}_{-0.22}$  & 7.71   & 7.72  \\                         
GS11     & 3 & $-$                     & $-$                      & 8.51 & 8.50 & 8.49 & 8.41 & 8.52   &  $-$ & 7.75 &  7.75  \\                        
HK500    & 4 & $-$                     & $-$                      & 8.57 & 8.54 & 8.39 & 8.35 & 8.49   &  $-$ & 7.82 & 7.78    \\                           
\hline                                                                                                                            
38       & 1 & $-$ 		       & $-$		          & 8.35 & 8.54 & 8.55 & 8.55 & 8.56   &  $-$ 	&  7.79  &  7.81 \\                           
39       & 1 &$-$                      & $-$                      & 8.47 & 8.56 & 8.56 & 8.54 & 8.55   &  $-$ 	&  7.86  &  7.87 \\                            
GS12     & 3 & $-$                     & $-$                      & 8.14 & 8.47 & 8.54 & 8.49 & 8.52   &  $-$   &  7.50  &  7.54  \\                        
HK453    & 4 & $-$                     & $-$                      & 8.21 & 8.48 & 8.47 & 8.49 & 8.52   &  $-$   &  7.54  &  7.54  \\                         
disc6    & 5 & $-$                     & $-$                      & 8.39 & 8.48 & 8.50 & 8.46 & 8.53   &  $-$   &  7.70  &  7.73  \\                       
\hline                                                                                                                            
47       & 1 & $-$ 		       & $-$		          & 8.56 & 8.53 & 8.53 & 8.44 & 8.51   &  $-$   &  7.79  &  7.81 \\                        
48       & 1 & $-$                     & $-$                      & 8.39 & 8.56 & 8.57 & 8.57 & 8.57   &  $-$   &  7.85  &  7.87  \\                       
GS13     & 3 & $-$                     & $-$                      & 8.52 & 8.57 & 8.58 & 8.54 & 8.54   &  $-$   &  7.92  &  7.93  \\                         
HK230    & 4 & $-$                     & $-$                      & 8.58 & 8.57 & 8.47 & 8.51 & 8.55   &  $-$   & 7.98   &  7.94  \\

\hline
\end{tabular}
\label{comparison}
References for the ID and line intensities: (1) this work, (2) \citet{Stan:10a},
(3) \citet{Bre:99a}, (4) \citet{Gar:87a}, (5) \citet{Patt:12a}.
\end{minipage}
\end{table*}

\section{Discussion}

The question of whether a single straight-line fit describes well the metallicity
gradient in a galaxy is often raised \citep[see e.g.][]{Patt:12a}. This does not concern 
us here. We have fitted straight lines in order to see the dependence of
the slope on the method used for the abundance determination and to measure the
dispersion of the results around these fits. We would get similar dispersions if we
just measured the dispersion in abundances for regions located at similar
galactocentric distances. Besides, the low dispersions around the gradient
shown by the abundances derived with the ONS, C, and N2 methods suggest that
straight-line fits are good first approximations to the data.

The main objective of this work is to study the effectiveness  of the methods
in producing robust measurements of abundance variations across a galaxy. One
assumption we make is that the more robust methods will produce lower dispersions
around the gradient. In the presence of azimuthal variations, we do not expect that
any method will imply a dispersion lower than the real one. We think that this is
a reasonable assumption. Hence, we use the dispersions introduced by the different
methods as a measure of their robustness or sensitivity to the observational data set used.
Note that the robustness of a method should not be confused with its reliability.
The more robust methods will not necessarily provide better results. The reliability
of the direct method depends on the validity of its assumptions; the reliability of
the strong-line methods depends on their calibration and on their application to objects
that are well represented in the calibration samples. In what follows, we will centre
our discussion in the robustness of the methods, and will assume that if a strong-line
method does not provide a good estimate of the oxygen abundance, it is
possible that it can be better calibrated to do so.

\begin{figure*}
\includegraphics[width=0.53\textwidth, trim=10 0 10 0, clip=yes, angle=90]{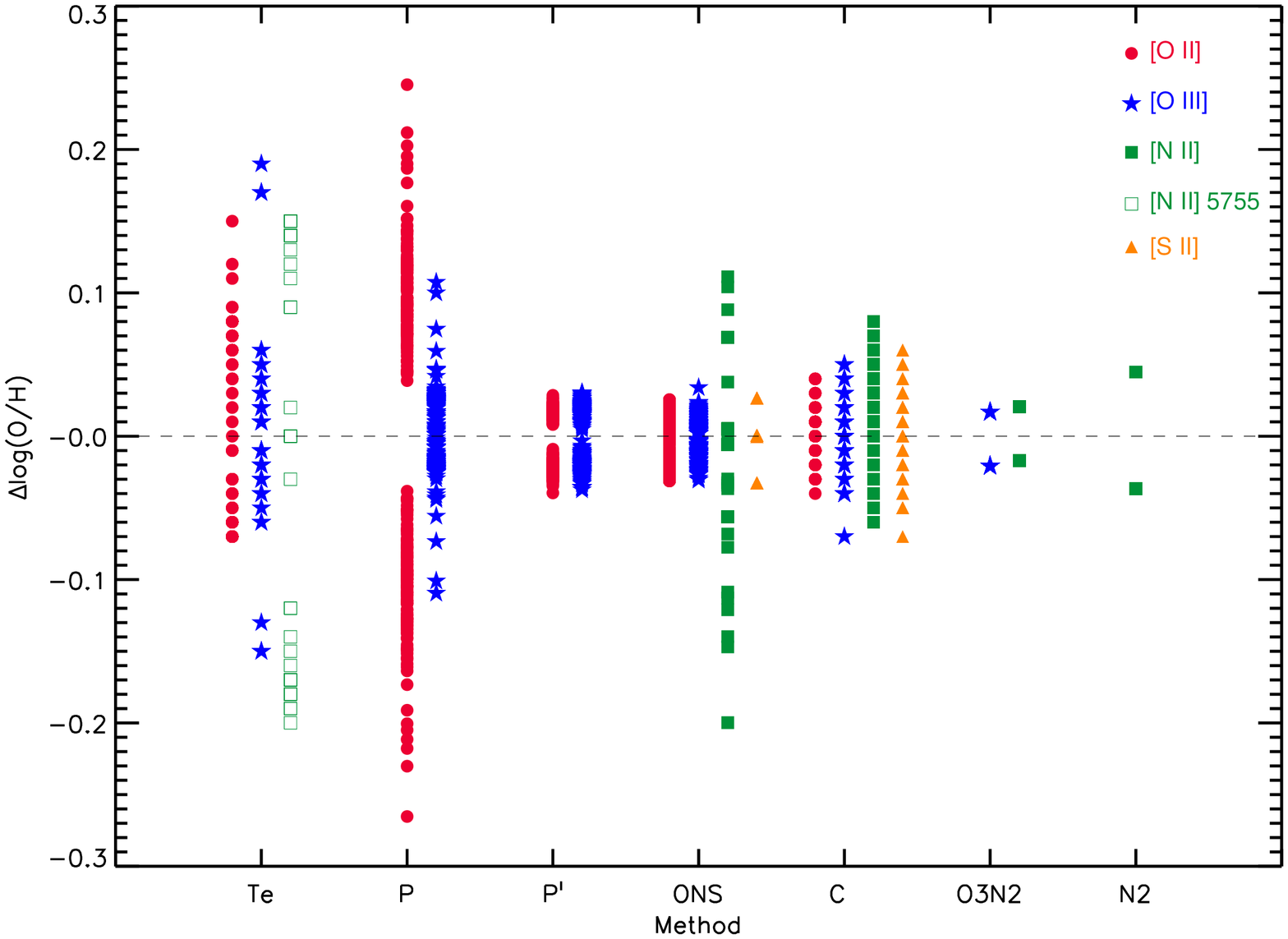}
\caption{Changes in the oxygen abundances for our sample of \ion{H}{ii} regions introduced
by changes of 20 per cent in the main line ratios used by each method. Circles, stars,
squares and triangles are used to represent changes in line ratios involving lines of
[\ion{O}{ii}], [\ion{O}{iii}], [\ion{N}{ii}], and [\ion{S}{ii}], respectively.
[\ion{N}{ii}] 5755' implies changes in the
[\ion{N}{ii}]~$(\lambda6548+\lambda6583)/\lambda5755$ intensity ratio, `[\ion{N}{ii}]' 
implies changes in the [\ion{N}{ii}]~$(\lambda6548+\lambda6583)/$H$\beta$ ratio for the ONS
method, and the [\ion{N}{ii}]~$\lambda6583/$H$\beta$ ratio for the C, O3N2, and N2
methods.} 
\label{sens}
\end{figure*}

\begin{figure}
\includegraphics[width=0.45\textwidth, trim=25 0 40 0, clip=yes]{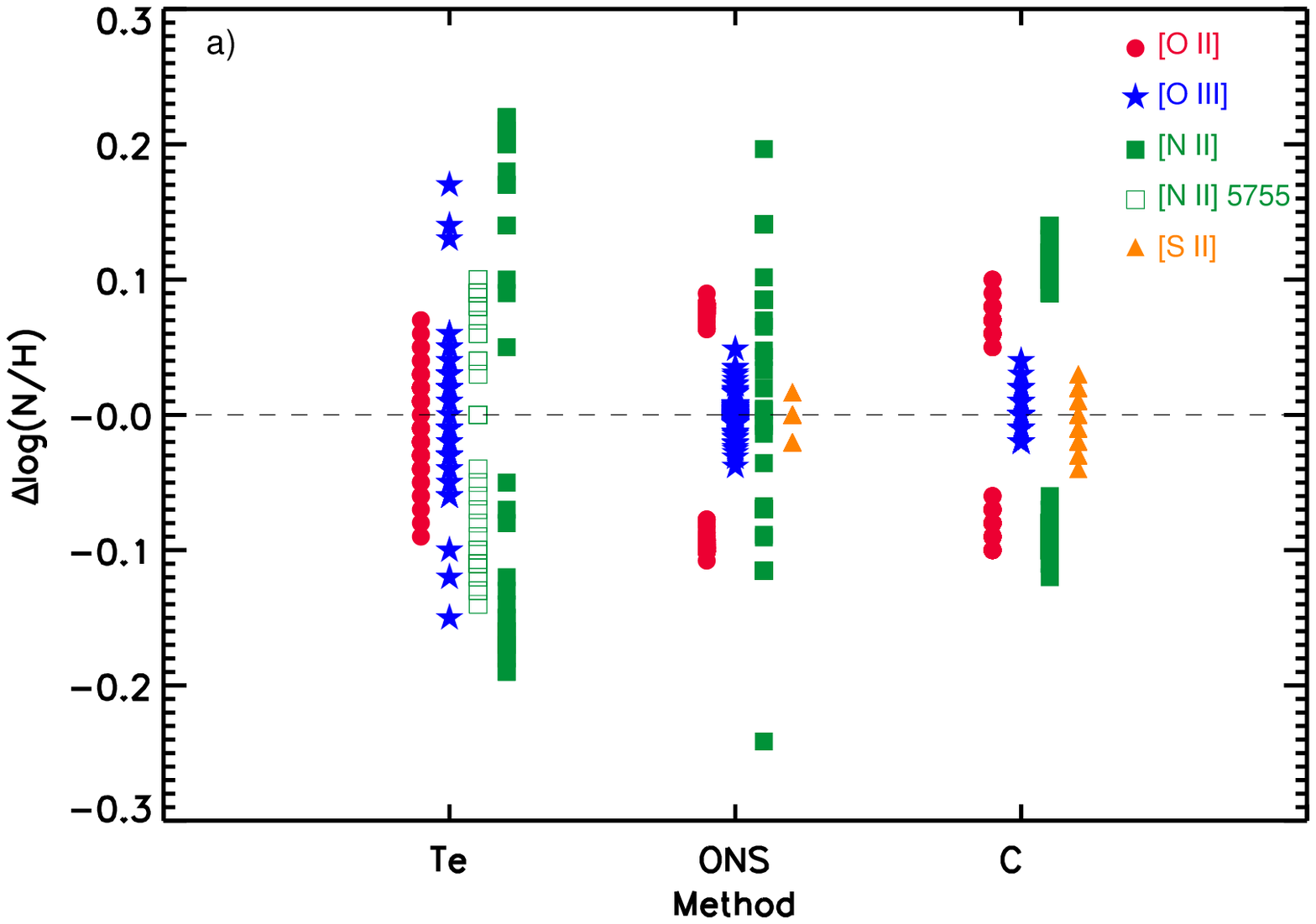}
\includegraphics[width=0.45\textwidth, trim=25 0 40 0, clip=yes]{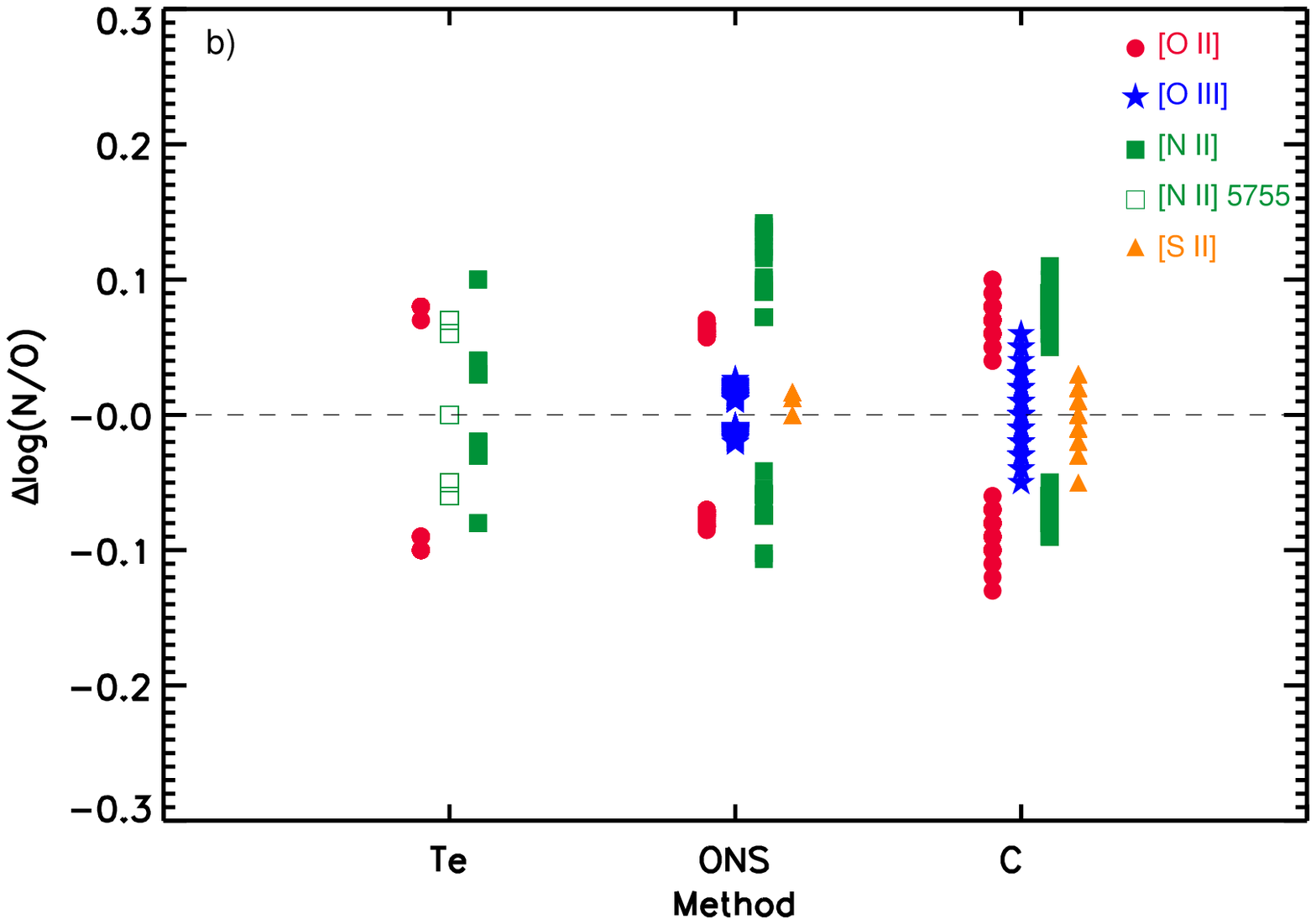}
\caption{Changes in the N/H and N/O abundance ratios for our sample of \ion{H}{ii} regions
introduced by changes of 20 per cent in the main line ratios used by each method. Circles,
stars, squares and triangles are used to represent changes in line ratios involving lines of
[\ion{O}{ii}], [\ion{O}{iii}], [\ion{N}{ii}], and [\ion{S}{ii}], respectively.
`[\ion{N}{ii}] 5755'
implies changes in the [\ion{N}{ii}]~$(\lambda6548+\lambda6583)/\lambda5755$ ratio for the
direct method, `[\ion{N}{ii}]'  implies changes in the
[\ion{N}{ii}]~$(\lambda6548+\lambda6583)/$H$\beta$ ratio for the direct method and the ONS
method, and the [\ion{N}{ii}]~$\lambda6583/$H$\beta$ ratio for the C method.} 
\label{sens-n}
\end{figure}

Figs.~\ref{sens} and \ref{sens-n} illustrate the sensitivity of each method to the main
line ratios involved in the calculations. We plot in these figures the changes in the O/H, N/H,
and N/O abundance ratios resulting from changes of 20 per cent in the main line intensity
ratios involved in the calculations for all the regions in our sample. Note that in these
figures `[\ion{N}{ii}] $\lambda$5755' identifies the results of changes in the
[\ion{N}{ii}]~$(\lambda6548+\lambda6583)/\lambda5755$ temperature diagnostic,
`[\ion{N}{ii}]' identifies the results of changes in the
[\ion{N}{ii}]~$(\lambda6548+\lambda6583)/$H$\beta$ ratio for the ONS method and the
direct method, and the [\ion{N}{ii}]~$\lambda6583/$H$\beta$ ratio for the C, O3N2 and
N2 methods.

As expected, the results of the direct method are very sensitive to
variations in the line ratio used to derive the electron temperature. This makes
this method vulnerable to different observational problems, especially the ones
arising from the measurement of the intensity of the weak line
[\ion{N}{ii}]~$\lambda5755$. The P method of \citet{Pil:05a} shows an even larger
sensitivity to changes in the line ratio [\ion{O}{ii}]~$\lambda3727/$H$\beta$,
making it vulnerable to problems introduced by atmospheric differential refraction
and defective flux calibrations or extinction corrections. This is even more clear if we
consider the dispersion from the gradient implied by this method when the regions where it has
problems are included in the fit, 0.24 dex. It can be argued that
these two line ratios, [\ion{N}{ii}]~$(\lambda6548+\lambda6583)/\lambda5755$ and
[\ion{O}{ii}]~$\lambda3727/$H$\beta$, are the ones most likely to be affected by observational
problems, making the direct method and the P method the least robust methods, in
agreement with our results. In fact, the dispersions around the gradients listed in
Tables~\ref{Results} and \ref{Results-nitrogen} can be qualitatively understood in terms of the
sensitivity of the methods to changes in these two line ratios, shown in Figs.~\ref{sens} and
\ref{sens-n}.

The results we obtain for N/O with the direct method, shown in Fig.~\ref{N-gradient}c,
can be used to illustrate this effect, since the N/O abundances derived with this method depend
mainly on the value of $T_{\rm e}$ implied by the
[\ion{N}{ii}]~$(\lambda6548+\lambda6583)/\lambda5755$ intensity ratio and on the
[\ion{N}{ii}]~$(\lambda6548+\lambda6583)/$[\ion{O}{ii}]~$\lambda3727$ intensity ratio. Our
observed regions (the diamonds in this figure) have larger N/O ratios than most of the regions
observed by \citet{Patt:12a} and \citet{Stan:10a}. The values we find for
$T_{\rm e}$([\ion{N}{ii}]) in our observed regions are generally higher than those we find for
the regions of \citet{Patt:12a} by an amount that can explain the differences in this abundance
ratio. On the other hand, we find similar values of $T_{\rm e}$([\ion{N}{ii}]) for our regions
and the regions observed by \citet{Stan:10a}. In this case the differences can be attributed
to the large values of the [\ion{O}{ii}]~$\lambda3727/$H$\beta$ measured by \citet{Stan:10a} in
several regions, which are higher than the ones observed by us and by \citet{Patt:12a}. If we
compare the values of this line ratio for the regions that have temperature determinations and
are located at galactocentric distances between 4 and 11 kpc, we find a range of values 149--338
for our observed objects and 229--327 for the regions observed by \citet{Patt:12a}, whereas the
regions observed by \citet{Stan:10a} span a range of 180--660. This translates into an
[\ion{N}{ii}] to [\ion{O}{ii}] line intensity ratio of 0.30--0.59 (this work), 0.15--0.45
\citep{Patt:12a}, and 0.16--0.29 \citep{Stan:10a}. The high values of $c(\mbox{H}\beta)$ found
by \citet{Stan:10a} contribute in part to these differences, but they are already present in
their observed intensities.

Any work whose objective is the determination of abundances in \ion{H}{ii} regions
considers an achievement the detection of the weak lines required for the
calculation of electron temperature, since temperature-based abundances are
expected to be more reliable than those based on strong-line methods. Our results
in Fig.~\ref{gradients} and Tables~\ref{Results} and \ref{comparison} suggest
otherwise. The abundances derived with the direct method are very sensitive to the
assumed temperature, which in turn is sensitive to the line intensity ratio used
for the diagnostic, as illustrated in Fig.~\ref{sens}. The precision required to
get a good estimate of this ratio is often underestimated.

The P method, based on the intensities of strong [\ion{O}{ii}] and [\ion{O}{iii}]
lines relative to H$\beta$, seems to be working slightly better in many cases,
although there are regions whose abundances show large deviations from their
expected values. The results shown in Fig.~\ref{sens} suggest that the spectra of
these regions might have problems related with atmospheric differential refraction,
flux calibration or extinction correction. In this context, it would be useful to
check whether the deviations are correlated with the airmass during the
observation, but none of the papers whose spectra we use provides the airmass values
of their observations. The new calibration of the P method of \citet{Pil:14a},
which we have called P$^\prime$ above, is less sensitive to the
[\ion{O}{ii}]~$\lambda3727/$H$\beta$ line ratio and performs much better when used
to derive the oxygen abundance gradient, implying a slope of $-0.008$ dex
kpc$^{-1}$ and a dispersion around the gradient of 0.09 dex. However, the
calibration sample of the P$^\prime$ method includes regions with abundances
determined using the C method. Since we have centred here on methods calibrated
with \ion{H}{ii} regions that have temperature measurements, we only show the
results of the P method in Fig.~\ref{gradients} and Table~\ref{Results}.

The other strong-line methods, especially the ONS, C, and N2 methods, seem to be
working remarkably well (see the dispersions in Table~\ref{Results} and
Fig.~\ref{gradients}). These methods suggest that azimuthal variations, if present,
are very small. The low dispersion implied by the N2 method is especially remarkable,
since it is due to a low dispersion in the values of the
[\ion{N}{ii}]~$\lambda\lambda6548,6583/$H$\alpha$ intensity ratio
that can only arise if N/H and the degree of ionization are both varying
smoothly across the disc of M81. Since these quantities and N/O might show different
variations in other environments, the N2 method will not necessarily give consistent
results for O/H when applied to \ion{H}{ii} regions in other galaxies or to regions
located near galactic centres. In fact, \citet{PMC:09} find that the N2 method can
lead  to values of O/H that differ from the ones derived with the direct method by up
to an order of magnitude.  The ONS and C methods should be preferred for this
reason, although we note that any strong-line method could easily fail for
\ion{H}{ii} regions whose properties are not represented in the calibration sample
\citep{Sta:10b}. The best estimates of the chemical abundances in \ion{H}{ii} regions
implied by forbidden lines will still be based on the measurement of electron
temperatures, but we stress that they require data of high quality.

This is illustrated by the work of \citet{Bre:11a}, who found that the scatter in
the oxygen abundances derived with the direct method in the central part of the
galaxy M33 is around 0.06 dex when using his observations, whereas the data of
\citet{Roso:08a} lead to much larger variations, with a dispersion of 0.21 dex. The
spectra of \citet{Bre:11a} were deeper than the ones observed by \citet{Roso:08a},
which might explain this result, although there could be other effects involved in
the explanation. Another example of the low dispersion that can be found with the
direct method is provided by \citet{Bre:09a} for NGC~300, where 28 \ion{H}{ii}
regions covering a relatively large range of galactocentric distances show a
dispersion around the gradient of only 0.05 dex.

\section{Summary and Conclusions}

We have used long slit spectra obtained with the GTC telescope to extract spectra
for 48 \ion{H}{ii} regions in the galaxy M81. We have added to this sample the
spectra of 68 \ion{H}{ii} regions in M81 observed by different authors
\citep{Gar:87a,Bre:99a,Stan:10a,Patt:12a}. This sample was re-analysed using 
the line intensities reported in each work. We followed the same procedure that we applied in
our sample to calculate physical properties, chemical abundances and galactocentric distances
for these \ion{H}{ii} regions.
The final sample contains 116 \ion{H}{ii} regions that cover a range of galactocentric distances of 3--33 kpc. We
have used these data to derive the oxygen and nitrogen abundance gradients in M81. We could
calculate the electron temperature and apply the direct method to 31 \ion{H}{ii}
regions of the sample. We used different strong-line methods to derive oxygen
and nitrogen abundances for the full sample. We have chosen strong-line methods
calibrated with large samples of \ion{H}{ii} regions with temperature-based abundance
determinations: the P method of \citet{Pil:05a}, the ONS method of \citet{Pil:10a},
the C method of \citet{Pil:12a}, and the O3N2 and N2 methods calibrated by
\citet{Marino:13a}.

We have fitted straight lines to the variation with galactocentric distance of the oxygen
abundances implied by each method. We find metallicity gradients with slopes that go from
$-0.010$ to $-0.002$ dex kpc$^{-1}$. The two extreme values are derived with the
P method and the direct method (the shallower value). These two methods are the ones that are
more sensitive to variations in two of the line ratios, most likely affected by observational
problems, [\ion{N}{ii}]~$(\lambda6548+\lambda6583)/\lambda5755$ and
[\ion{O}{ii}]~$\lambda3727/$H$\beta$, and show the largest dispersions around the gradient,
0.25 and 0.15 dex, respectively, whereas the ONS, C, O3N2, and N2 methods imply
oxygen abundance gradients in the range from $-0.008$ to $-0.006$ dex kpc$^{-1}$ and very low
dispersions, equal to 0.06 dex, for the C and N2 methods, 0.07 dex for the
ONS method, and 0.09 dex for the O3N2 method.  Since we are using observations from five
different works, which are likely to be affected by diverse observational problems by differing
amounts, we argue that this implies that the ONS, C, and N2 methods are the more robust
methods. Our comparison of the results implied by the different methods for several of our
objects that were also observed by other authors agree with this result. The low dispersions
also imply that if there are azimuthal variations in the oxygen abundance in M81, they must be
small.

In the case of N/H, we have used the direct method, the C method, and the ONS method,
and find gradients of $-0.025$ to $-0.011$ dex kpc$^{-1}$, with the direct method providing
again the shallower slope and the largest dispersion around the fit, 0.18 dex, versus 0.15 dex
for the ONS method and 0.12 dex for the C method. For N/O we find
slopes that go from $-0.020$ to $-0.008$ dex kpc$^{-1}$, with the latter value derived with the
direct method, although for this abundance ratio the dispersions are similar for the three
methods, 0.11--0.13 dex. The dispersions around the gradients obtained with the different
methods for O/H, N/H, and N/O can be qualitatively
accounted for by considering the sensitivity of the methods to the two critical line ratios,
[\ion{N}{ii}]~$(\lambda6548+\lambda6583)/\lambda5755$ (our main temperature diagnostic in this
work) and [\ion{O}{ii}]~$\lambda3727/$H$\beta$.

All the robust methods use the intensity of [\ion{N}{ii}]~$\lambda6584$, and the
N2 method is only based on the intensity of this line with respect to H$\alpha$.
Since nitrogen and oxygen do not vary in lockstep because they are produced by
different types of stars, and their relative abundances depend on the star
formation history of the observed galactic region \citep[see, e.g.,][]{Mol:06}, the
low dispersions around the oxygen abundance gradient found with the robust methods
suggest that both N/O and the degree of ionization vary smoothly along the disc of
M81. On the other hand, the different values of N/O generally found for regions
with similar oxygen abundances imply that strong-line methods that use the
intensities of [\ion{N}{ii}] lines will produce different oxygen abundances in
regions that have similar values of O/H but different values of N/O. The ONS and
C methods, that use line ratios involving several ions, and also estimate 
the N/H abundance ratio, can be expected to correct for this
effect, at least for regions whose properties are well represented in their
calibration samples, but the N2 method by itself cannot achieve this correction.
Since our analysis indicates that the available observations do not allow
reliable determinations of abundances through the direct method in this galaxy,
and since we do not know if the more robust methods are working properly for the
observed \ion{H}{ii} regions, the magnitude of the metallicity gradient in M81 remains
uncertain. These issues should be further investigated using observations
of \ion{H}{ii} regions in different environments that allow the determination of
electron temperatures and N and O abundances through the direct method. The large
dispersion in the abundances around the gradient that we find here when using the
direct method implies that these observations should have high quality in order to
get meaningful results. For the time being, we recommend the use of the ONS or
C methods when no temperature determinations are possible or when the available
determinations are of poor quality.

\section*{Acknowledgements}
We thank the anonymous referee for useful comments that helped us to improve
the content of the paper. Based on observations made with the GTC, installed in the Spanish Observatorio del Roque de los Muchachos of the Instituto
de Astrof\'isica de Canarias, in the island of La Palma. We acknowledge support
from Mexican CONACYT grants CB-2010-01-155142-G3 (PI: YDM), 
CB-2011-01-167281-F3 (PI: DRG) and CB-2014-240562 (PI: MR). 
K.Z.A.-C. acknowledges support from CONACYT  grant 351585.


\appendix
\bsp
\label{lastpage}
\end{document}